%% file: Main_FinalVersion.tex
\documentclass[journal]{IEEEtran}

\ifCLASSINFOpdf
\else
\fi

\usepackage{xcolor}
 \usepackage{tikz}
    \usetikzlibrary{positioning}
    \usetikzlibrary{calc}
    \usetikzlibrary{shapes}
    \usetikzlibrary{patterns} 
\usepackage{amsmath,amssymb,amsthm}
\usepackage{pgfplots}
    \pgfplotsset{compat=1.15}
\usepackage{cite}

\usepackage{soul}

\usepackage{adjustbox}

\definecolor{myDarkGreen}{rgb}{0.00000,0.48824,0.00000}%

\def\lf{\left\lfloor}
\def\rf{\right\rfloor}

\def\define{\stackrel{\Delta}{=}}

\def\calA{\mathcal{A}}

\def\kmax{K^\bullet}

\def\emax{E^\bullet}

\tikzstyle{point-visible} = [debugpoint,inner sep=0pt, minimum size = 2,fill=red] 
\tikzstyle{point-invisible} = [coordinate]
\tikzstyle{point} = [point-invisible] 

\colorlet{amp3color}{green!70!black}%
\colorlet{amp1color}{blue!70!black}%
\colorlet{amp5color}{red!70!black}%
\colorlet{amp7color}{orange!90!black}%
\colorlet{ESScolor}{green!70!black}
\colorlet{SMcolor}{red!70!black}
\colorlet{CCDMcolor}{red}
\colorlet{MPDMcolor}{blue}
\colorlet{SSHcolor}{black}

\def\Ptx{P_{\text{tx}}}

\begin{document}

\title{Kurtosis-limited Sphere Shaping for Nonlinear Interference Noise Reduction in Optical Channels}

\author{Yunus~Can~G\"{u}ltekin,~\IEEEmembership{Member,~IEEE,}
Alex~Alvarado,~\IEEEmembership{Senior~Member,~IEEE,}
Olga~Vassilieva,~\IEEEmembership{Senior Member,~IEEE,}
Inwoong~Kim,~\IEEEmembership{Senior Member,~IEEE,}
Paparao~Palacharla,~\IEEEmembership{Senior Member,~IEEE,}
Chigo~M.~Okonkwo,~\IEEEmembership{Senior~Member,~IEEE}
and~Frans~M.~J.~Willems,~\IEEEmembership{Life Fellow,~IEEE}
\thanks{Manuscript received June 23, 2021; revised September 10, 2021 and October 11, 2021; accepted October 13, 2021. 
The work of Y. C. G\"{u}ltekin and A. Alvarado has received funding from the European Research Council (ERC) under the European Union's Horizon 2020 research and innovation programme (grant agreement No 757791).
This article was presented in part at the European Conference on Optical Communication (ECOC), Bordeaux, France, September 2021. \emph{(Corresponding author: Yunus Can G\"{u}ltekin.)}}
\thanks{Y. C. G\"{u}ltekin, A. Alvarado and F. M. J. Willems are with the Information and Communication Theory Lab, Signal Processing Systems Group, Department of Electrical Engineering, Eindhoven University of Technology, Eindhoven 5600 MB, The Netherlands (e-mails: \{y.c.g.gultekin, a.alvarado, f.m.j.willems\}@tue.nl).}
\thanks{C. M. Okonkwo is with the Electro-Optical Communications Group, Institute for Photonics Integration, Department of Electrical Engineering, Eindhoven University of Technology, Eindhoven 5600 MB, The Netherlands (e-mail: cokonkwo@tue.nl).}
\thanks{O. Vassilieva, I. Kim and P. Palacharla are with the Fujitsu Network Communications Inc., Richardson, 75082 TX, USA (e-mails: \{olga.vassilieva, inwoong.kim, paparao.palacharla\}@us.fujitsu.com).}
}

\markboth{Draft, \today}{}

\maketitle

\begin{abstract}
Nonlinear interference (NLI) generated during the propagation of an optical waveform through the fiber depends on the fourth order standardized moment of the channel input distribution, also known as kurtosis.
Probabilistically-shaped inputs optimized for the linear Gaussian channel have a Gaussian-like distribution with high kurtosis.
For optical channels, this leads to an increase in NLI power and consequently, a decrease in effective signal-to-noise ratio (SNR).
In this work, we propose kurtosis-limited enumerative sphere shaping (K-ESS) as an algorithm to generate low-kurtosis shaped inputs.
Numerical simulations at a shaping blocklength of 108 amplitudes demonstrate that with K-ESS, it is possible to increase the effective SNRs by 0.4 dB in a single-span single-channel scenario at 400 Gbit/s.
K-ESS offers also a twofold decrease in frame error rate with respect to Gaussian-channel-optimal sphere shaping.
\end{abstract}


\IEEEpeerreviewmaketitle

\section{Introduction}\label{sec:intro}
\IEEEPARstart{T}{he} capacity of the linear additive white Gaussian noise (AWGN) channel can be achieved by transmitting independent and identically distributed (i.i.d.) Gaussian inputs~\cite[Ch. 9]{CoverT2006_ElementsofInfoTheo}. 
Recently, probabilistic amplitude shaping (PAS) was proposed as a coded modulation strategy that achieves this capacity~\cite{BochererSS2015_ProbAmpShap}. 
PAS combines an outer amplitude shaper with an inner channel encoder.
The amplitude shaper selects the amplitudes of the inputs while the encoder selects their signs. 
For the AWGN channel, amplitude shapers are designed such that the resulting input distribution has a Gaussian-like behavior~\cite{SchulteB2016_CCDM,Fehenberger2019_MPDM,Schulte2019_SMDM,GultekinHKW2019_ESSforShortWlessComm,Fehenberger2020_HCSSviaLUT_letter}.
Motivated by the performance improvements obtained for the AWGN channel, PAS has attracted considerable attention in optical communication systems. 
It was demonstrated both numerically and experimentally that PAS with AWGN-optimal shapers provides significant gains in reach or in data rate for long-haul optical links~\cite{FehenbergerABH2016_OnPSofQAMforNLFC_short,Buchali2016_RateAdaptReachIncrease_letter,idler2017_fieldtrialPS_letter,Goossens2019_FirstExperimentESS_letter,Amari2019_ESSreachincrease_letter,Amari2019_IntroducingESSoptics_letter}. 

The propagation of the optical field through the fiber is subject to nonlinearities.
Its propagation is governed by the nonlinear Shr\"{o}dinger equation.
During this propagation, single-channel transmission experiences intra-channel nonlinear interference (NLI), while multi-channel transmission (via wavelength-division multiplexing, WDM) also experiences inter-channel NLI.
Often, detection schemes designed for linear channels are used for optical communications, and hence, the NLI is implicitly treated as noise.
The so-called Gaussian noise (GN) model approximately describes the variance of this NLI independently of the modulation format and its probability distribution~\cite{Poggiolini2014_GNmodel}.
This approach has been shown to be inaccurate, especially for short links, and thus, the enhanced GN (EGN) model was later proposed in~\cite{Dar2013_PropertiesNLIN,Carena2014_EGNmodel,Dar2014_AccumultionNLIN}.
The EGN model predicts that the fourth order standardized moment (i.e., the {\it kurtosis}) of the input distribution affects the amount of NLI: inputs with higher kurtosis lead to higher NLI, and hence, lower effective signal-to-noise ratios (SNRs)~\cite[Sec. IV-E1]{FehenbergerABH2016_OnPSofQAMforNLFC_short}.

PAS framework was considered for long-haul WDM optical communications for the first time in~\cite{Fehenberger2015_LDPCcodedPS}.
However, only the Maxwell-Boltzmann (sampled Gaussian, MB) distribution is considered, and the SNRs are computed using the GN model.
In~\cite{FehenbergerABH2016_OnPSofQAMforNLFC_short}, the EGN model is adopted, and the kurtosis-dependence of the NLI is used to explain the reason why AWGN-optimal shaping strategies have an SNR-penalty with respect to uniform signaling: the Gaussian distribution has a relatively high kurtosis.
Results in~\cite{Renner2017_ExperimentalPScomparison_letter} experimentally confirmed this kurtosis-dependence for single-span links with 9 and 80 WDM channels.
The results in~\cite{Renner2017_ExperimentalPScomparison_letter} also show that gains obtained by optimizing inputs for optical channels exist but are limited.

In~\cite{Pan2016_PS16QAMinWDM}, the input distribution is optimized for a 16-ary constellation using the EGN model for transmission with 9 WDM channels and assuming the NLI is circularly symmetric AWGN. 
It is concluded that AWGN-optimal inputs are good enough.
However, there are two caveats:
First, as found in~\cite{Dar2014_AccumultionNLIN} and stated in~\cite{Pan2016_PS16QAMinWDM}, the kurtosis has the greatest influence for short links. 
On the other hand in~\cite[Fig. 2]{Pan2016_PS16QAMinWDM}, it is not possible to evaluate the effect of kurtosis for short distances.
In this figure, all the considered schemes have mutual information (MI) converging to $\log_216=4$~bits per symbol when the link consists of fewer than 15 spans of 80 km, i.e., the optimum distribution converges to the uniform distribution.
Therefore, it is not possible to draw a conclusion about shaping for short links.
Second, the MI is estimated assuming that the NLI noise is circularly symmetric Gaussian.
However, for short distances (especially for single-span), this is not the case~\cite[Fig. 1]{Dar2013_PropertiesNLIN}.

In~\cite{Sillekens2018_nonlinearitytailoredPS_letter}, optimized input distributions are computed considering kurtosis, and gains in achievable information rates (AIRs)---although small---are observed for 256- and 1024-ary constellations for 5 WDM channels.
Unlike~\cite{Pan2016_PS16QAMinWDM}, a single-span 200 km link is considered in which the effect of kurtosis is expected to be significant and the NLI noise is not circularly symmetric Gaussian.
However, the study is limited to AIR computations, and no shaping algorithm is implemented.

In~\cite{Yoshida2020_HiDMmasive}, a shaping technique, namely hierarchical distribution matching~\cite{YoshidaKA2019_HIDM,Civelli2020_HiDM}, is discussed.
The implementation of this technique is based on lookup tables (LUTs).
In~\cite[Sec. IV]{Yoshida2020_HiDMmasive}, these LUTs are generated such that channel input sequences that have higher $j^{\text{th}}$ order moments are discarded for $j \leq 8$. 
It is concluded that for systems where the dominant source of impairment is the noise generated by the amplifiers, considering only the second order moment is optimum.
However, for short links, optimizing higher order moments improves the performance~\cite[Fig. 4]{Yoshida2020_HiDMmasive}.

A modified version of the MB distributions in which higher order moments can be controlled is introduced in~\cite{Tehrani2018_supergaussian} as ``super-Gaussian'' distributions.
These distributions are shown to provide higher nonlinear tolerance than MB distributions based on i.i.d. symbol-wise shaping.
Later, these super-Gaussian distributions have been implemented in a chip in~\cite{Napoli2020_800GASIC}.
Recently in~\cite{Hansen2020_NonGaussianNoise}, an optimization of the channel input distribution is performed for transmission with 7 WDM channels over nonlinear fiber channels with non-Gaussian noise.
Simulations again based on i.i.d. symbol-wise shaping demonstrate reach increase for dispersion-managed fibers with these optimized distributions.
A concise review of some of the other works attempting to use non-AWGN-optimal distributions to decrease NLI can be found in~\cite[Sec. I]{Renner2017_ExperimentalPScomparison_letter}.

\begin{figure}[t]
    \centering
	\resizebox{0.89\columnwidth}{!}{\includegraphics{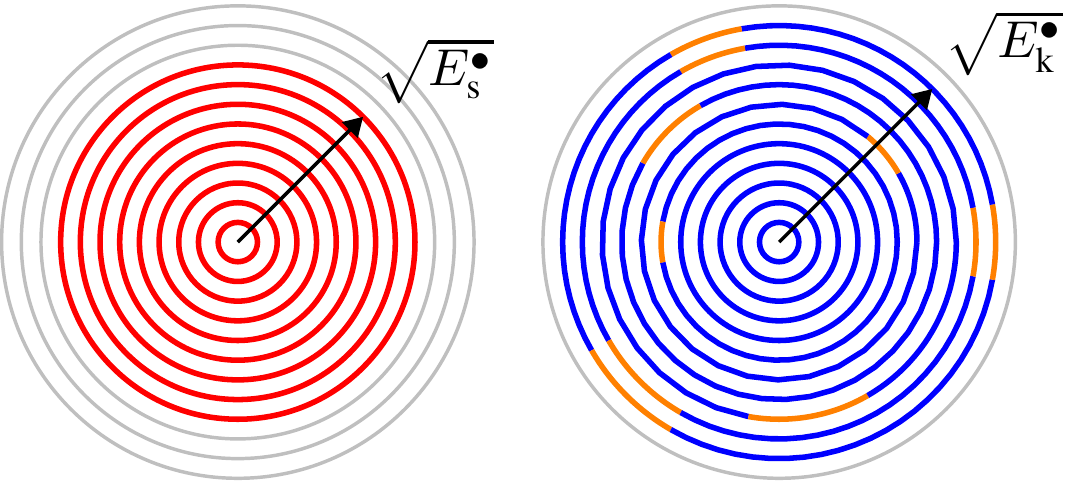}}
    \caption{Illustration of the shell occupation in an $N$-dimensional sphere: (Left, red) sphere shaping, (right, blue) kurtosis-limited sphere shaping. 
Due to the limit $\kmax$ on kurtosis, some sequences (orange) in the sphere are left out of the shaping set. 
Since a higher energy usually implies a higher kurtosis, the sequences that are excluded due to $\kmax$ are mostly from outermost shells.
To compensate for the decrease in number of sequences, i.e., in rate, that results from the $\kmax$ constraint, additional energy shells are included, i.e., $\emax_{\text{k}} > \emax_{\text{s}}$.
Similar illustrations were provided in~\cite[Fig. 1]{GultekinHKW2019_ESSforShortWlessComm},~\cite[Fig. 6]{Gultekin2019Arxiv_Comparison}, and~\cite[Fig. 2]{Fehenberger2020_HCSSviaLUT_letter}.} 
    \label{fig:sphericalillustration}
\end{figure}

Although it is well recognized in the literature that kurtosis is important in NLI generation, no constructive algorithms exist implementing specifically kurtosis-limited shaping, other than the algorithms that can target any distribution, e.g., constant composition (CCDM)~\cite{SchulteB2016_CCDM} or multiset-partition distribution matching (MPDM)~\cite{Fehenberger2019_MPDM}, etc.
Here, we attempt to provide such an algorithm using an ``indirect'' method according to the terminology of~\cite{Calderbank1990_NonEquipSignaling}: by changing the bounding geometry of the signal space, rather than trying to obtain a target input distribution.
Spherical signal structures (as illustrated in ~\ref{fig:sphericalillustration} (left)) are optimum for the AWGN channel~\cite{Gultekin2018_ConstShapforShortBlocks}.
However, other bounding geometries may lead to more efficient signal sets for non-AWGN channels as we discussed in~\cite[Sec. I]{Gultekin2019Arxiv_Comparison}.
With this motivation, in this work:
\begin{enumerate}
    \item 
    We introduce a new amplitude shaping approach which we call kurtosis-limited sphere shaping (KLSS).
    KLSS extends the idea of sphere shaping, which is to construct a spherical signal space to obtain inputs with Gaussian-like distribution, by imposing a constraint on the kurtosis of the signal points.
    With this additional constraint, we effectively propose to use a non-spherical signal structure for optical channels as illustrated in Fig.~\ref{fig:sphericalillustration} (right).
    
    \item
    We show that KLSS generates signal sets with slightly higher average energy than that of sphere shaping, but with a smaller average kurtosis.
    
    \item
    We implement a KLSS-based amplitude shaper using a modified version of the enumerative sphere shaping (ESS) algorithm~\cite{GultekinWHS2018_ApproxEnumerative,GultekinHKW2019_ESSforShortWlessComm}.
    This new algorithm, i.e., K-ESS, creates an invertible mapping from binary indices to amplitude sequences within the KLSS set.
\end{enumerate}
Simulation results show that PAS with K-ESS recovers some of the SNR-penalty observed with AWGN-optimal ESS for single-span links.
End-to-end decoding results demonstrate that with KLSS, smaller frame error rates (FERs) than sphere shaping and uniform signaling are obtained for single-span links.
It is also observed that the additional gain provided by K-ESS over ESS decreases as the number of WDM channels or the number of spans increase.
ESS can be recovered as a special case of K-ESS, and thus, K-ESS generalizes ESS.
Comparisons of ESS, i.e., sphere shaping, with i.i.d. symbol-wise shaping and with other finite-blocklength shaping architectures such as CCDM and MPDM can be found in~\cite{Skvortcov2020_NLtolerantLUTshaping_letter,Skvortcov2021_HCSSforExtendedReachSingleSpanLins_letter} and in~\cite{Goossens2019_FirstExperimentESS_letter,Amari2019_ESSreachincrease_letter,Amari2019_IntroducingESSoptics_letter,FehenbergerA2019_AnalyzeOptimizeDMforNLchan}, respectively.

The remainder of the paper is organized as follows. In Sec.~\ref{sec:systemmodel}, some background information is provided, and the system model is described.
In Sec.~\ref{sec:klss}, KLSS and K-ESS are introduced.
Numeric results are then discussed in Sec.~\ref{sec:numeric}.
And finally, some conclusions are drawn in Sec.~\ref{sec:conclusion}.

\section{System Model \& Problem Statement}\label{sec:systemmodel}
\subsection{Probabilistic Amplitude Shaping}
In this study, we restrict our attention to PAS~\cite{BochererSS2015_ProbAmpShap}.
PAS is a layered coded modulation strategy where in the first layer, an amplitude shaper determines the amplitudes $a^N = (a_1, a_2,\cdots, a_N)$ of the channel inputs.
In the second layer, a forward error correction (FEC) code is used to encode these amplitudes to obtain the corresponding signs $s^N\in\{-1,+1\}^N$ in the form of parity bits.
Typically, this is achieved by using a systematic FEC code.
Assuming that the FEC encoding produces uniform signs~\cite[Sec. IV-A2]{BochererSS2015_ProbAmpShap}, the channel input distribution and the properties of the channel input sequences are determined by the amplitude shaper.

\subsection{Amplitude Shaping for the AWGN Channel}
The capacity of the average-power-constrained AWGN channel can be achieved when the input $X$ is a zero-mean i.i.d. Gaussian.
Corresponding input sequences $x^N = (x_1, x_2,\dotsc, x_N)$ can be shown to be confined in an $N$-sphere for $N$ large enough.
Reciprocally, it is possible to show that if the signal space is bounded by an $N$-sphere, lower-dimensional input distributions converge to a Gaussian for $N$ large enough.
Motivated by this duality, there are two main approaches to realize amplitude shaping in the PAS framework for the AWGN channel.
The first one is to try to obtain an MB distribution at the output of the amplitude shaper by using distribution matching.
Distribution matching can be based on CCDM~\cite{SchulteB2016_CCDM}, MPDM~\cite{Fehenberger2019_MPDM}, etc.
The second approach is to try to obtain a spherically-constrained signal space by using sphere shaping, e.g., via shell mapping~\cite{Schulte2019_SMDM}, enumerative sphere shaping (ESS)~\cite{GultekinHKW2019_ESSforShortWlessComm}, Huffman-coded sphere shaping (HCSS)~\cite{Fehenberger2020_HCSSviaLUT_letter}, etc.
In this work, we focus on the second approach via ESS and refer the reader to~\cite{Gultekin2019Arxiv_Comparison} for a comparison of distribution matching and sphere shaping.
Sphere shaping considers the energy-constrained amplitude sequences in the set
\begin{equation}
\calA^\bullet = \left\{a^N  : \sum_{i=1}^{N} a_i^2 \leq \emax \right\}, \label{eq:ElimitedSet}
\end{equation}
where $a_i \in \calA = \{1, 3,\cdots,M-1\}$ for $i = 1, 2,\dotsc, N$, and $M=2^m$ for a positive integer $m$.
The set $\calA$ is the amplitude alphabet of $M$-ary amplitude-shift keying ($M$-ASK) and of the real and imaginary parts of $M^2$-ary quadrature amplitude modulation ($M^2$-QAM).
There are $|\calA^\bullet|$ amplitude sequences in $\calA^\bullet$, and the input length of an amplitude shaper that outputs $a^N\in\calA^\bullet$ is defined as $k = \lf\log_2 |\calA^\bullet|\rf$~bits.
In \eqref{eq:ElimitedSet}, $\emax$ is the maximum energy that the signal points are allowed to have, i.e., the squared radius of the $N$-sphere.
By changing the value of $\emax$, the shaping rate $k/N$ can be adjusted.

\subsection{Amplitude Shaping for the Optical Channel}
In~\cite{Poggiolini2014_GNmodel}, the effective SNR of an optical signal after propagation is defined as $\text{SNR}_{\text{eff}} \define \Ptx / (\sigma^2_{\text{ASE}}+\sigma^2_{\text{NLI}})$ where $\Ptx$ is the optical launch power, $\sigma^2_{\text{ASE}}$ is the variance of the noise introduced by the amplifiers, and $\sigma^2_{\text{NLI}}$ is the variance of the NLI.
In the GN model of~\cite{Poggiolini2014_GNmodel}, the NLI is modeled as a circularly symmetric Gaussian noise whose properties are assumed to be independent of the input modulation format and distribution.
This assumption was later shown to be inaccurate, and the EGN model was proposed~\cite{Dar2013_PropertiesNLIN,Carena2014_EGNmodel}.

The $k^{\text{th}}$ order standardized moment is defined as~\cite[eq. (19)]{FehenbergerABH2016_OnPSofQAMforNLFC_short}
\begin{equation}
\mu_k = \frac{E[|X-E[X]|^k]}{(E[|X-E[X]|^2])^{\frac{k}{2}}} = E[|X|^k], \label{eq:highermoementbootom}
\end{equation}
for a channel input $X$ that is symmetric around the origin and has unit energy.
The EGN model shows that the variance $\sigma^2_{\text{NLI}}$ of NLI can be expressed as~\cite[eq. (17)]{FehenbergerABH2016_OnPSofQAMforNLFC_short}
\begin{equation}
    \sigma^2_{\text{NLI}} = \Ptx^3 \left[ \chi_0 + \left( \mu_4-2\right)\chi_4 + \left(\mu_4-2\right)^2\chi{'}_4 + \mu_6\chi_6 \right]. \label{eq:nlivariance}
\end{equation}
In \eqref{eq:nlivariance}, $\chi_0, \chi_4, \chi{'}_4,$ and $\chi_6$ are coefficients that are determined by fiber parameters.
Considering that $\chi{'}_4$ is typically negative and that $\mu_4$ is typically smaller than 2, \eqref{eq:nlivariance} implies that as $\mu_4$ or $\mu_6$ increases, the NLI variance increases as well~\cite{FehenbergerABH2016_OnPSofQAMforNLFC_short}.
Consequently, the SNR-penalty observed for AWGN-optimal input distributions in~\cite[Sec. IV-E1]{FehenbergerABH2016_OnPSofQAMforNLFC_short} is explained by the higher kurtosis values of such distributions~\cite[Table II]{FehenbergerABH2016_OnPSofQAMforNLFC_short}.

\begin{figure}[t]
\centering
\resizebox{\columnwidth}{!}{\includegraphics{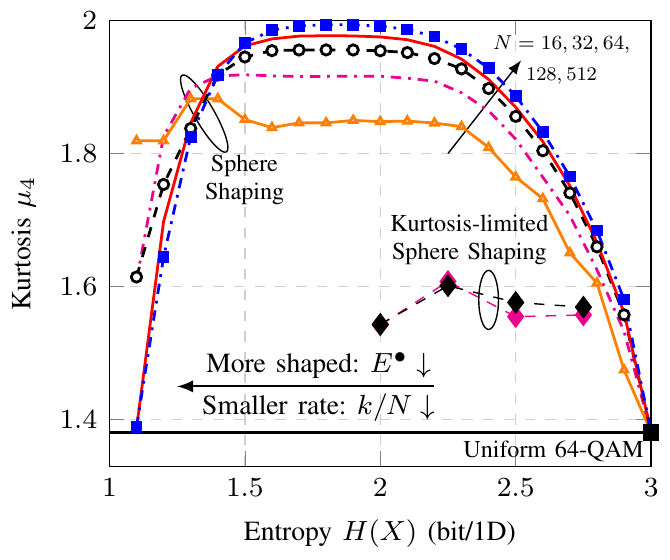}}
\caption{Kurtosis $\mu_4$ of the channel inputs obtained using sphere shaping with 64-QAM. Black square corresponds to uniform 64-QAM with kurtosis 1.381~\cite[Table II]{FehenbergerABH2016_OnPSofQAMforNLFC_short}. For uniform phase-shift keying constellations, kurtosis is equal to 1. 
The pink and the black diamonds correspond to KLSS with $N=32$ and $64$, respectively.
}
\label{fig:EssKurtosis}
\end{figure}

As an example in Fig.~\ref{fig:EssKurtosis}, we plotted the kurtosis of the channel input $X$ with amplitudes selected with sphere shaping.
As the distribution is shaped more, i.e., as $\emax$, $k/N$ and the entropy of $X$ all decrease, kurtosis increases and then decreases for small rates.
Furthermore, when the shaping redundancy is around 1 bit, the kurtosis is very close to that of a continuous Gaussian. 
We conclude from Fig.~\ref{fig:EssKurtosis} that finite-blocklength sphere shaping leads to input distributions with high kurtosis which increases NLI variance, and accordingly, decreases effective SNR. 
However, the increase in AIRs due to shaping is larger, and a positive net gain is obtained, especially for long-haul transmission~\cite[Sec. IV-E]{FehenbergerABH2016_OnPSofQAMforNLFC_short}.
In the next section, we will propose a new sphere shaping approach to lower the kurtosis, and thus, recover the SNR-penalty.

\section{Kurtosis-limited Sphere Shaping}\label{sec:klss}
\subsection{Kurtosis Constraint on Amplitude Sequences}
We want to devise a shaping technique that has smaller kurtosis than that of sphere shaping.
Consider the energy- and kurtosis-constrained amplitude sequences in the set
\begin{equation}
\calA^\blacktriangle = \left\{a^N  : \sum_{i=1}^N a_i^2 \leq \emax \mbox{ and } \sum_{i=1}^N a_i^4 \leq \kmax  \right\}. \label{eq:EKlimitedSet}
\end{equation} 
We call this set the kurtosis-limited sphere shaping (KLSS) set.
Similar to sphere shaping, the input length of an amplitude shaper that outputs $a^N\in\calA^\blacktriangle$ is defined as $k = \lf\log_2|\calA^\blacktriangle|\rf$~bits. 
The shaping rate $k/N$ can be adjusted by changing the values of $\emax$ and $\kmax$.
As $\kmax$ decreases, sequences with high kurtosis are eliminated from the shaping set, and hence, the rate decreases.
As $\kmax\rightarrow\infty$, $\calA^\blacktriangle$ converges to the regular sphere shaping set $\calA^\bullet$ in \eqref{eq:ElimitedSet}.

\begin{figure}[t]
\centering
\resizebox{\columnwidth}{!}{\input{EmaxKmaxPlane_Histograms.tikz}}
\caption{Histograms of kurtosises of sequences for $N=64$ with 64-QAM. 
Kurtosis of a sequence $a^N$ is computed using \eqref{eq:highermoementbootom} with the corresponding empirical distribution.
{\bf Inset:} The $\emax\kmax$-plane with $k/N=1.00$, $1.25$ and $1.50$~bit/amplitude shaping rate contours. Here, the black square effectively represents uniform signaling where there is no constraint on energy or kurtosis, i.e., the signal space is effectively bounded by an $N$-cube.}
\label{fig:histogramKurtosis}
\end{figure}
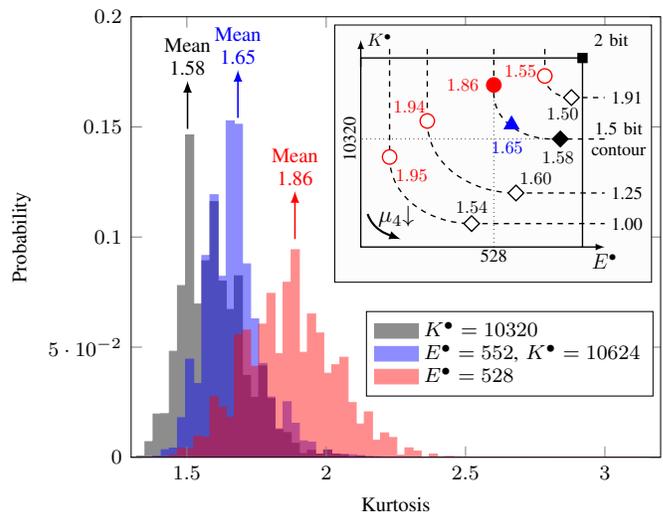

The effect of changing $\emax$ and/or $\kmax$ can be visualized as shown in the inset figure of Fig.~\ref{fig:histogramKurtosis} where we show the $\emax\kmax$-plane.
For each point on this plane, the corresponding set $\calA^\blacktriangle$ in \eqref{eq:EKlimitedSet} can be found.
The filled red circle corresponds to sphere shaping with only an energy constraint, i.e., $\emax=528$.
On the other hand, the filled black diamond represents the case with only a kurtosis constraint, i.e., $\kmax=10320$.
The curve that connects these points is the contour line of constant shaping rate $k/N=1.50$~bits per amplitude (bit/amplitude). 
On this curve, as we move away from sphere shaping, kurtosis decreases, however average energy increases.
We have also illustrated the contour lines of 1.91, 1.25, and 1.00 bit/amplitude.
One important observation here is that the maximum possible decrease in kurtosis, i.e., the difference between $\mu_4$'s of the red circle and the black diamond on a given contour line, is very small (below $0.05$) for shaping rates above 1.9 bit/amplitude.
This is expected since as the shaping rate increases, the number of possible $(\emax,\kmax)$ pairs decreases, with every corresponding $\calA^\blacktriangle$ converging to a uniform signal set, i.e., an $N$-cube.
In the extreme case where the shaping rate is $m-1$ bit/amplitude, there is a single shaping set---which is not shaped anymore---and a single possible kurtosis value.
We note that this extreme, i.e., uniform signaling, is represented by the filled black squares in Figures~\ref{fig:EssKurtosis} and~\ref{fig:histogramKurtosis}.

In Fig.~\ref{fig:EssKurtosis}, the minimum $\mu_4$ value that can be obtained with KLSS is also shown (diamond markers).
We see that the kurtosis limitation indeed leads to smaller $\mu_4$ than that of sphere shaping at the same rate.
Furthermore, $\mu_4$ is relatively flat and roughly independent of $N$ for KLSS.

Figure~\ref{fig:histogramKurtosis} also demonstrates the effect of the kurtosis constraint $\kmax$ in \eqref{eq:EKlimitedSet}.
Three cases are considered:
Red histogram belongs to sphere shaping, i.e., only $\emax$ is finite and the signal space is as illustrated in Fig.~\ref{fig:sphericalillustration} (left).
Blue histogram belongs to KLSS, i.e., both $\emax$ and $\kmax$ are finite, and the signal space is as illustrated in Fig.~\ref{fig:sphericalillustration} (right).
Black histogram belongs to the other extreme where only $\kmax$ is finite, i.e., there is no constraint on the energy of the signal points.
These three histograms correspond to the filled red circle, blue triangle, and black diamond in the inset figure, respectively.
The trend to have more sequences with smaller kurtosis as $\kmax$ decreases is self-evident in Fig.~\ref{fig:histogramKurtosis}.

\begin{figure}[t]
\centering
\resizebox{\columnwidth}{!}{\includegraphics{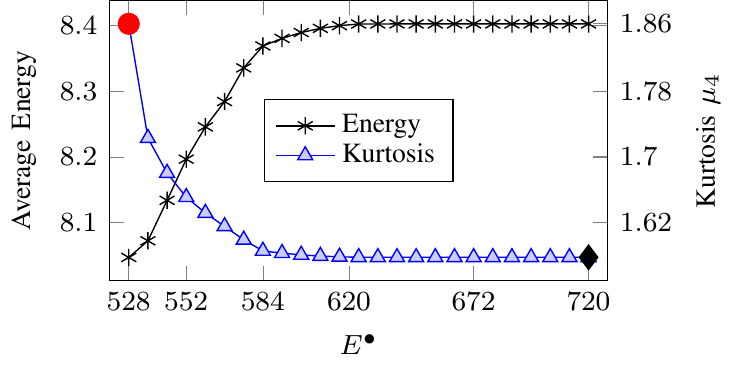}}
\caption{Trade-off between the average energy per symbol and kurtosis $\mu_4= E[|X|^4]$ for the pairs of $\emax$ and $\kmax$ on the 1.5 bit/amplitude shaping rate contour in the inset of Fig.~\ref{fig:histogramKurtosis}. 
The black diamond corresponds to the kurtosis-limited scheme in Fig.~\ref{fig:EssKurtosis} and in the inset of Fig.~\ref{fig:histogramKurtosis}. The red circle corresponds to the energy-limited scheme in the inset of Fig.~\ref{fig:histogramKurtosis}.}
\label{fig:ShapGainKurtosis}
\end{figure}

Figure~\ref{fig:ShapGainKurtosis} shows the trade-off between average energy $\sum_{a\in\calA}p(a)a^2$ and kurtosis when moving on a constant shaping rate contour.
As $\emax$ increases, the average energy per symbol increases. 
At the same time, $\kmax$ decreases leading to a decrease in kurtosis.
We also see from Fig.~\ref{fig:ShapGainKurtosis} that after approximately $\emax=584$ ($\kmax=10320$), increasing $\emax$ only has marginal effect.
This is because the proportion of sequences that have energy larger than 584 and kurtosis smaller than 10320 that are included in the shaping set is extremely small.

\subsection{Enumerative Sphere Shaping}
\begin{figure}[t]
    \centering
\resizebox{\columnwidth}{!}{\includegraphics{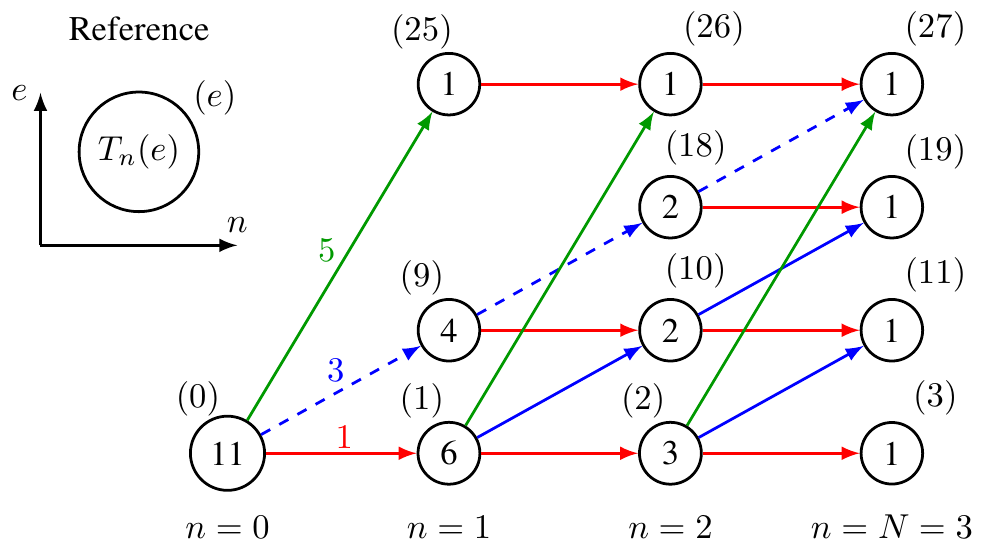}}
    \caption{Bounded-energy enumerative amplitude trellis for $\calA = \{1, 3, 5\}$, $N = 3$ and $\emax = 28$.}
    \label{fig:ess_trellis}
\end{figure}

Enumerative sphere shaping (ESS) is an algorithm to index amplitude sequences in an $N$-sphere $a^N\in\calA^\bullet$~\cite{GultekinWHS2018_ApproxEnumerative,GultekinHKW2019_ESSforShortWlessComm}.
To realize ESS, first an enumerative amplitude trellis is generated, as shown in Fig.~\ref{fig:ess_trellis} for $N=3$, $\calA=\{1, 3, 5\}$ and $\emax=27$.
Here, nodes in the $n^{\text{th}}$ column represent the accumulated energy of the sequences for their first $n$ amplitudes, more precisely, $\sum_{i=1}^n a_i^2$.
The nodes are labeled with the energy level $(e)$ that they represent.
Color-coded branches that connect a node in $(n-1)^{\text{th}}$ column to a node in $n^{\text{th}}$ column represent $a_n$.
Each $3$-amplitude sequence is shown by a $3$-branch path that starts at node $(0)$ and ends in a node in the final (rightmost) column.
Possible values for sequence energy are $3$, $11$, $19$ and $27$.
As an example, the path that corresponds to $(3, 3, 3)$ with energy $27$ is drawn with dashed lines in Fig.~\ref{fig:ess_trellis}.
The values written inside each node are computed using
\begin{equation}
T_n(e)\define \sum_{\substack{a \in \calA}}  T_{n+1}(e+a^2), \label{eq:backwardtrellis_ess}
\end{equation}
for $n = 0, 1,\dotsc, N-1$ where the initialization is as follows:
\begin{equation}
T_N(e) = \left\{
     \begin{array}{lr}
       1 :& e \leq \emax,\\
       0 :& \mbox{ otherwise }.
     \end{array}
   \right. \label{trellisEq_ess}
\end{equation}
In \eqref{eq:backwardtrellis_ess}, $T_n(e)$ is the number of ways to reach a final node starting from the node of energy $e$ in the $n^{\text{th}}$ column. 
Thus, $T_0(0)$ gives the number of sequences $|\calA^\bullet|$ represented in the trellis, which is $11$ in Fig.~\ref{fig:ess_trellis}.
This means there are $11$ 3-amplitude sequences with energy no greater than $27$.
Note that only the accumulated energy values that can occur are considered in Fig.~\ref{fig:ess_trellis}.
Based on this enumerative trellis, ESS realizes a lexicographical mapping from indices, i.e., $\lf \log_2 T_0(0) \rf=k$-bit strings, to amplitude sequences in $\calA^\bullet$~\cite[Sec. III-C]{GultekinHKW2019_ESSforShortWlessComm}.

\subsection{KLSS Implementation Based on ESS}
\begin{figure*}[t]
    \centering
\resizebox{\textwidth}{!}{\includegraphics{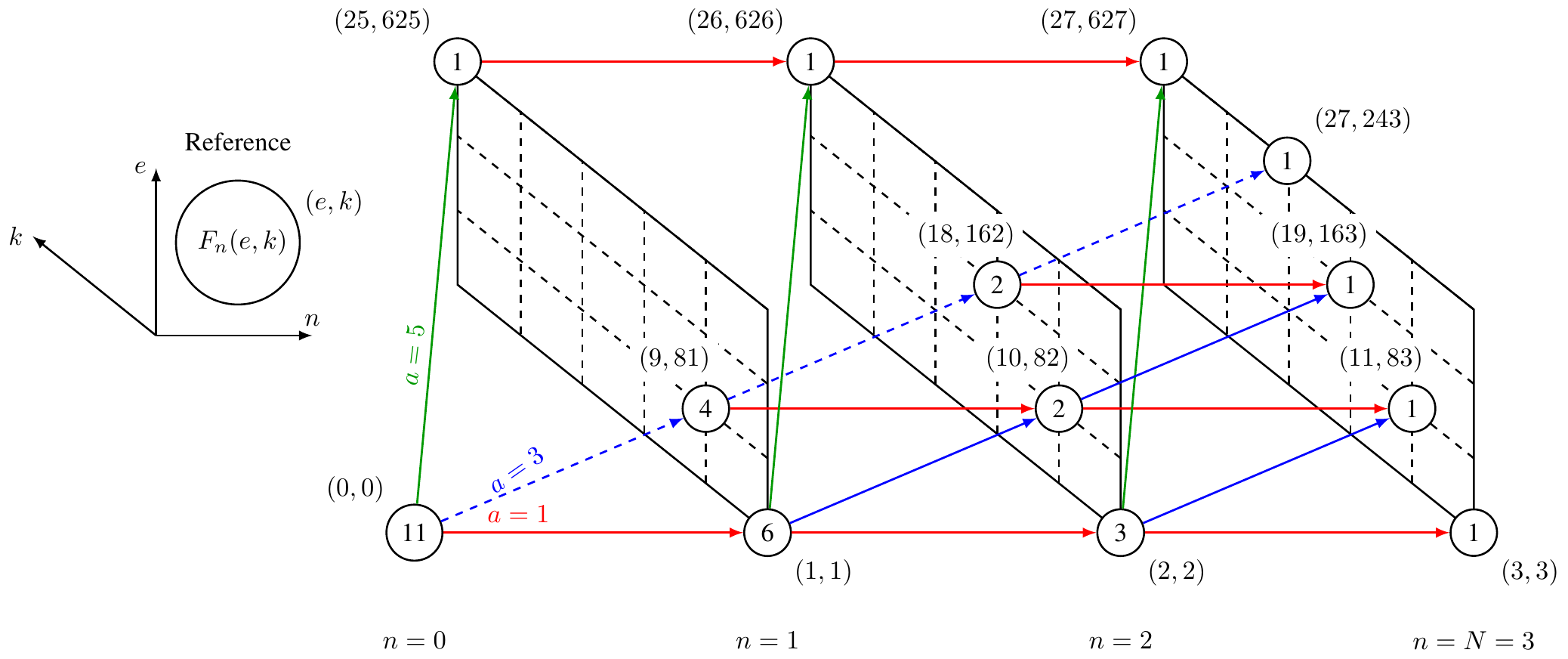}}
    \caption{Bounded-energy and bounded-kurtosis enumerative amplitude trellis for $\calA = \{1, 3, 5\}$, $N = 3$, $\emax = 28$ and $\kmax=627$.}
    \label{fig:kess_trellis}
\end{figure*}

To realize kurtosis-limited sphere shaping (KLSS), we introduce a modified version of the enumerative amplitude trellis explained in the previous section.
We again use an example trellis as shown in Fig.~\ref{fig:kess_trellis} for the same set of parameters as Fig.~\ref{fig:ess_trellis} and $\kmax=627$.
Now, nodes in the $n^{\text{th}}$ plane represent the accumulated energy and the accumulated kurtosis $\sum_{i=1}^n a_i^4$ of the sequences for their first $n$ amplitudes.
The nodes are labeled with the energy-kurtosis pair $(e, k)$ that they represent.
Similar to Fig.~\ref{fig:ess_trellis}, each $3$-amplitude sequence is shown by a $3$-branch path that starts at node $(0, 0)$ and ends in a node in the final (rightmost) plane.
Possible values for sequence energy-kurtosis pairs are $(3,3)$, $(11,83)$, $(19,163)$, $(27, 243)$ and $(27, 627)$.
Again, the path that corresponds to $(3, 3, 3)$ with energy $27$ and kurtosis $243$ is drawn with dashed lines in Fig.~\ref{fig:kess_trellis}.

Similar to \eqref{eq:backwardtrellis_ess}, the values written inside each node are computed using
\begin{equation}
F_n(e,k)\define \sum_{\substack{a \in \calA}}  F_{n+1}(e+a^2, k+a^4), \label{eq:backwardtrellis}
\end{equation}
for $n = 0, 1,\dotsc, N-1$  where the initialization is as follows:
\begin{equation}
F_N(e,k) = \left\{
     \begin{array}{lr}
       1 :& e \leq \emax \mbox{ and } k\leq \kmax,\\
       0 :& \mbox{ otherwise }.
     \end{array}
   \right. \label{trellisEq}
\end{equation}
In \eqref{eq:backwardtrellis}, $F_n(e,k)$ is the number of ways to reach a final node starting from the node of energy $e$ and kurtosis $k$ in the $n^{\text{th}}$ plane.
Accordingly, $F_0(0,0)=|\calA^\blacktriangle|$ is the number of sequences represented in the trellis.
Note that when you look to the trellis in Fig.~\ref{fig:kess_trellis} from the $en$-plane, the trellis in Fig.~\ref{fig:ess_trellis} is seen.
The difference is that now, we have control also over the maximum value of kurtosis allowed in the trellis.
As an example, in the maximum-energy-bounded trellis of Fig.~\ref{fig:ess_trellis}, when the final node of energy $27$ is included, all sequences with energy $27$ are in the shaping set: $(1, 1, 5)$, $(1, 5, 1)$, $(3, 3, 3)$ and $(5, 1, 1)$.
However, in the maximum-energy- and maximum-kurtosis-bounded trellis of Fig.~\ref{fig:kess_trellis}, it is possible to include the final node $(27,243)$ while excluding $(27,627)$.
This way, $(1, 1, 5)$, $(1, 5, 1)$ and $(5, 1, 1)$ can be left out, while $(3, 3, 3)$ stays in, which leads to a decrease in $\mu_4$.

Finally, the enumerative encoding and decoding algorithms explained in~\cite[Sec. III-C]{GultekinHKW2019_ESSforShortWlessComm} can be used also based on $F(e,k)$ to create an invertible mapping from $\lfloor \log_2 F(0,0) \rfloor=k$-bit indices to sequences in $\calA^\blacktriangle$.
We call this kurtosis-limited ESS (K-ESS), and we note that low-complexity implementation techniques introduced for ESS in~\cite{Gultekin2021_LowComplexityESS} can be applied to K-ESS straightforwardly.
It is important to note that in \eqref{trellisEq}, we only consider the pairs of $(e,k)$ in the $N^{\text{th}}$ (final) plane that can possibly occur for an $N$-tuple of amplitudes.
We determine these pairs of energy and kurtosis by an exhaustive search over all possible compositions of amplitudes.
This is different than ESS in which the corresponding initialization is done by setting $T_N^e=1$ for all $e\leq\emax$~\cite[Sec. III-B]{GultekinHKW2019_ESSforShortWlessComm}.

In the next section, we will simulate PAS with K-ESS to check whether we can recover the SNR-penalty observed for AWGN-optimal channel inputs due to their high kurtosis.

\section{Numeric Results}\label{sec:numeric}
\subsection{Simulation Setup}\label{ssec:simulationsetup}
To demonstrate the effectiveness of KLSS, we simulated single-channel transmission over a single span of standard single-mode optical fiber (SSMF) with an attenuation of 0.2 dB/km, a dispersion parameter of 17 ps/nm/km, and a nonlinear parameter of 1.3 1/W/km, followed by an erbium-doped amplifier with a noise figure of 5 dB.
The propagation of the optical field over the SSMF is simulated using the split-step Fourier method based on the nonlinear Shr\"{o}dinger and Manakov equations for single- and dual-polarized (SP and DP) transmission, resp.

The transmitter generates an SP or DP 50 GBd signal (with a root-raised cosine pulse with 10\% roll-off factor) using 64-QAM, yielding raw data rates of 300 and 600 Gbit/s, resp.
The transmission rate is 4 bit/2D symbol, yielding net data rates of 200 and 400 Gbit/s, resp.
With uniform signaling, this rate is obtained using a rate 2/3 channel code.
With PAS, we use a rate 5/6 channel code with an amplitude shaper with rate 1.5 bit/amplitude.
How these shaping and coding rates lead to a transmission rate of 2 bit/1D symbol is explained in~\cite[Footnote 1]{Gultekin2021_CompareOptimizeESS}.
The channel codes used are the low-density parity-check (LDPC) codes of the IEEE 802.11 standard with a codeword length of 648 bits~\cite{IEEE80211_2016_letter}.
For FER calculations, a ``frame'' is defined as an FEC block.
We consider both ESS and K-ESS with $N=108$.
There is a single ESS trellis that satisfies the 1.5 bit/amplitude shaping rate constraint, with $\emax=860$.
On the other hand, there are 42 $(\emax,\kmax)$ pairs that lead to a KLSS set~\eqref{eq:EKlimitedSet} with a shaping rate of 1.5 bit/amplitude.
Finally, for the mapping of coded bits to QAM symbols, the binary reflected Gray code is used~\cite[Defn. 2.10]{Szczecinski2015_BICMbook}.

\subsection{Mapping Amplitudes to 4D Symbols}
For DP transmission, two streams of QAM symbols should be generated to create 4D channel inputs, i.e., we need four amplitudes per channel use considering the in-phase and quadrature components of these streams.
Shaped amplitudes can be mapped to these four real dimensions in three different ways.
Following the nomenclature used in~\cite[Sec. II-C]{Skvortcov2021_HCSSforExtendedReachSingleSpanLins_letter}, we call these the {\it 1D, 2D and 4D symbol mapping} strategies.
In 1D symbol mapping, amplitudes of each real dimension are taken from an independent shaped sequence as shown in~\cite[Fig. 3(a)]{Skvortcov2021_HCSSforExtendedReachSingleSpanLins_letter}.
In 2D symbol mapping, amplitude of each polarization are taken from an independent shaped sequence as shown in~\cite[Fig. 3(b)]{Skvortcov2021_HCSSforExtendedReachSingleSpanLins_letter}.
In 4D symbol mapping, amplitudes of the 4D channel inputs are take from the same shaped sequence as shown in~\cite[Fig. 3(c)]{Skvortcov2021_HCSSforExtendedReachSingleSpanLins_letter}.
We note that in the SP case, 1D and 2D symbol mappings correspond to the inter- and intra-shaper pairings described in~\cite[Sec. II-B]{Fehenberger2019_NLFiberInteractions_letter}, respectively.

\begin{figure}[t]
\centering
\resizebox{\columnwidth}{!}{\includegraphics{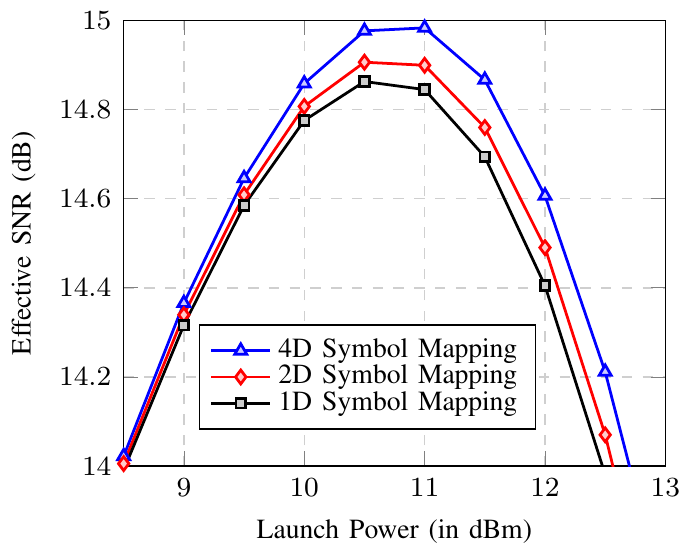}}
\caption{Effective SNR vs. launch power for the transmission of 400 Gbit/s DP-64-QAM over a 205-km-long SSMF.
The channel inputs are generated using PAS with K-ESS using the three symbol mapping strategies discussed in~\cite[Sec. II-C]{Skvortcov2021_HCSSforExtendedReachSingleSpanLins_letter}.}
\label{fig:effectofpairings}
\end{figure}

In Fig.~\ref{fig:effectofpairings}, the effective SNR is shown as a function of the launch power for K-ESS of 64-QAM with $N=108$, $\emax=1156$ and $\kmax=16556$ with different amplitude-to-symbol mapping strategies.
We see that as observed in~\cite[Fig. 12]{Skvortcov2021_HCSSforExtendedReachSingleSpanLins_letter}, 4D symbol mapping, i.e., grouping four consecutive amplitudes from a single shaped codeword into a 4D channel input symbol, maximizes the SNR around the optimal launch power.
Accordingly, we use the 4D mapping for the simulations discussed in the rest of the paper.

\subsection{Effective SNR Results}
\begin{figure}[t]
\centering
\resizebox{\columnwidth}{!}{\includegraphics{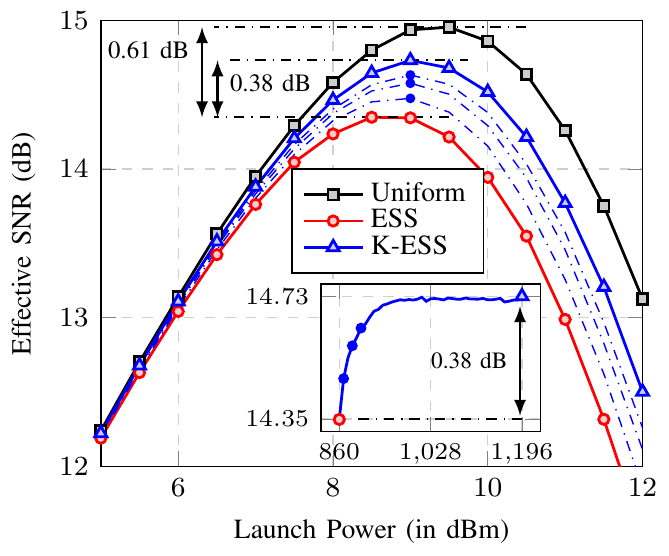}}
\caption{Effective SNR vs. launch power for the transmission of 200 Gbit/s SP-64-QAM over a 213-km-long SSMF.
{\bf Inset:} Effective SNR at the optimum launch power as a function of $\emax$ of the K-ESS trellis.
Optimum power is 9 dBm for all considered K-ESS schemes. 
Red circle represents regular sphere shaping.
Blue circles represent the K-ESS trellises for which the launch power vs. effective SNR performance is shown with the dash-dotted curves in the outer figure.
}
\label{fig:effsnrsingpol200}
\end{figure}

\begin{figure}[t]
\centering
\resizebox{\columnwidth}{!}{\includegraphics{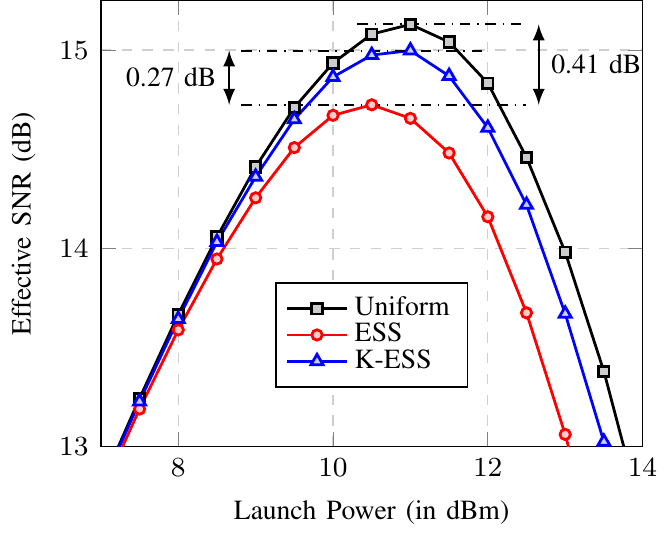}}
\caption{Effective SNR vs. launch power for the transmission of 400 Gbit/s DP-64-QAM over a 205-km-long SSMF.}
\label{fig:SNR_DP_400G}
\end{figure}

Figures~\ref{fig:effsnrsingpol200} and~\ref{fig:SNR_DP_400G} show the effective SNR as a function of optical launch power for SP and DP transmission, resp.
We see that due to its high kurtosis, ESS has an SNR-penalty with respect to uniform signaling.
The penalty is slightly higher in the case of SP transmission, i.e., 0.61 dB instead of 0.41 dB.
This is in agreement with~\cite[Sec. 5]{Dar2013_PropertiesNLIN} where it is discussed that the effect of kurtosis on NLI variance is slightly more significant for SP transmission.
Figures~\ref{fig:effsnrsingpol200} and~\ref{fig:SNR_DP_400G} show that K-ESS recovers some of this SNR-loss thanks to its smaller kurtosis.
Another observation from the effective SNR curves is that the optimum launch power for K-ESS is larger than that of ESS for both SP and DP transmission, and is the same as that of uniform signaling in the DP case.
This also implies that K-ESS can recover some of the nonlinearity tolerance that is lost when ESS is employed instead of uniform transmission.

In Fig.~\ref{fig:effsnrsingpol200}, we show the effective SNR for 4 out of the 42 available K-ESS trellises at the considered rate, i.e., as the operating point moves on the 1.5 bit shaping rate contour away from regular sphere shaping, similar to the inset figure of Fig.~\ref{fig:histogramKurtosis}.
This demonstrates that as $\emax$ increases and $\kmax$ decreases, the decrease in kurtosis $\mu_4$ results in an improvement in effective SNR.
In the inset figure, this trend is shown by plotting the effective SNR at the optimum launch power\footnote{The ``optimum launch power'' is the value at which the corresponding effective SNR or FER curve has its maximum.} as a function of $\emax$.
We see that as $\emax$ increases and $\mu_4$ decreases, effective SNR increases and converges to 14.73 dB.
This behavior is consistent with the convergence observed for average energy and kurtosis in Fig.~\ref{fig:ShapGainKurtosis}.\footnote{It is found in~\cite{Dar2013_PropertiesNLIN,Carena2014_EGNmodel} that the sixth order moment also plays a (relatively insignificant) role in NLI generation. For the sake of completeness, we report here that the sixth order moment $\mu_6$ also exhibits a convergence behavior for KLSS, similar to the one that is shown in Fig.~\ref{fig:ShapGainKurtosis} for kurtosis.}
The outermost K-ESS curve in Fig.~\ref{fig:effsnrsingpol200} (triangular markers) has the maximum effective SNR and is considered to be the ``best'' trellis for this setup.
We note that the ``best'' trellis may change depending on the parameters of the optical link and the communication scenario such as the number of channels and the bandwidth.
Furthermore, for the same setup, the trellis that maximizes the SNR and the one that minimizes the FER might differ. 
In the other figures presented in this paper, we therefore only plot the curve that belongs to the K-ESS trellis that optimizes the corresponding performance metric.
This optimum trellis is found by creating all K-ESS trellises and simulating their performance.

\subsection{End-to-end Decoding Results}
\begin{figure}[t]
\centering
\resizebox{\columnwidth}{!}{\includegraphics{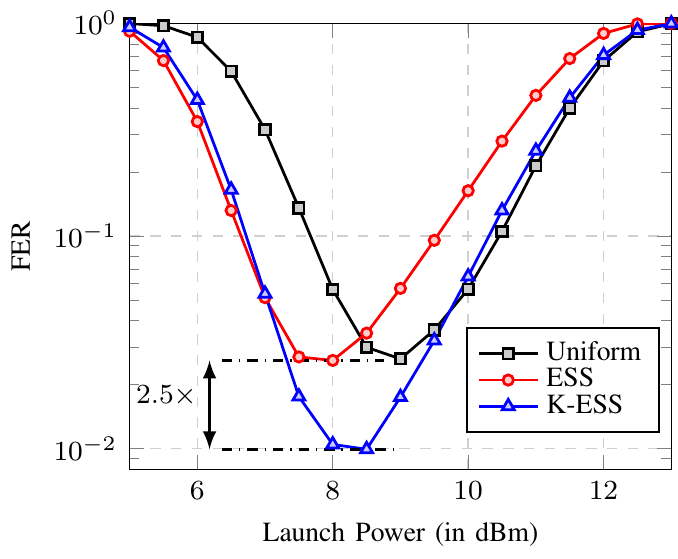}}
\caption{FER vs. launch power for the transmission of 200 Gbit/s SP-64-QAM over a 213-km-long SSMF.}
\label{fig:FER_SP_200G}
\end{figure}

\begin{figure}[t]
\centering
\resizebox{\columnwidth}{!}{\includegraphics{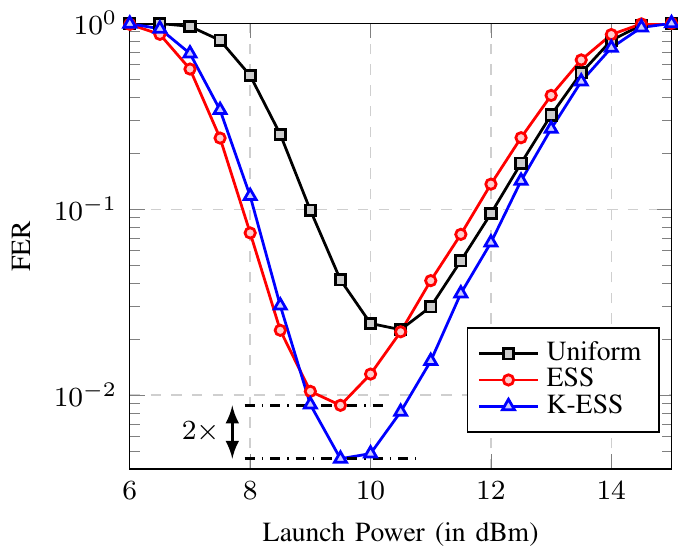}}
\caption{FER vs. launch power for the transmission of 400 Gbit/s DP-64-QAM over a 205-km-long SSMF.}
\label{fig:FER_DP_400G}
\end{figure} 

Figures~\ref{fig:FER_SP_200G} and~\ref{fig:FER_DP_400G} show the FER as a function of optical launch power for SP and DP transmission, respectively.
We see that although ESS had an SNR-penalty because of its high kurtosis with respect to uniform signaling (see Figs.~\ref{fig:effsnrsingpol200} and~\ref{fig:SNR_DP_400G}), the net gain (combined with the improvement in AIRs that results from shaping) in FERs is positive for DP transmission.
This is as observed in~\cite[Sec. IV-E]{FehenbergerABH2016_OnPSofQAMforNLFC_short}.
For SP transmission, the SNR-penalty and shaping gain cancel each other out.
In both SP and DP cases, ESS achieves the minimum FER at a smaller optimum launch power than uniform signaling.
This indicates that ESS has a decreased nonlinearity tolerance, but still, a decoding performance either as good as uniform signaling (SP) or better (DP).
With K-ESS, it is possible to obtain a better position at this trade-off.
We see that K-ESS further decreases the FERs by at least twofold with respect to ESS while increasing the optimum launch power.
This is thanks to the optimized trade-off enabled by K-ESS between the nonlinearity tolerance (due to decreased kurtosis) and the shaping gain in AIR.

\subsection{Rate Adaptivity}\label{ssec:rateadapt}
To investigate the performance of K-ESS at different transmission rates, we also simulated single-channel transmission over a single span of 227 km SSMF at a rate of 6 bit/4D symbol.
The transmitter generates a DP 50 GBd signal again using 64-QAM leading to a 300 Gbit/s net data rate.
With uniform signaling, this rate is obtained using a rate 1/2 channel code. 
With PAS, we again use a rate 5/6 channel code, now with an amplitude shaper with rate 1 bit/amplitude.
We consider both ESS and K-ESS with $N=108$.
The ESS trellis that satisfies 1 bit/amplitude rate is with $\emax=428$.
However there are 9 $(\emax,\kmax)$ pairs that lead to a KLSS set as in \eqref{eq:EKlimitedSet} with rate 1 bit/amplitude.

\begin{figure}[t]
\centering
\resizebox{\columnwidth}{!}{\includegraphics{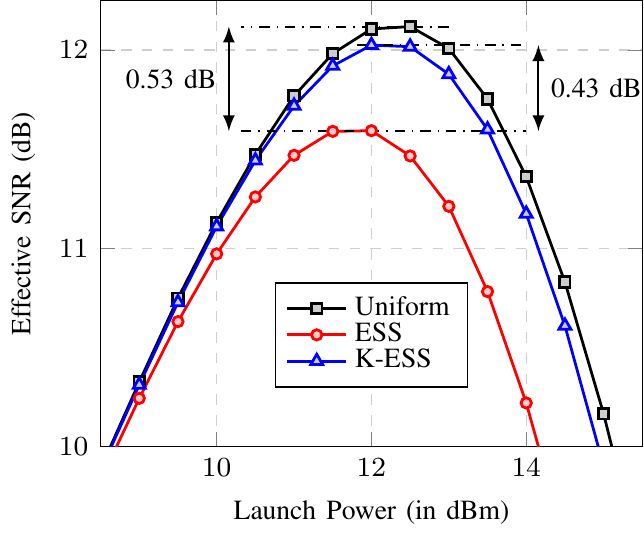}}
\caption{Effective SNR vs. launch power for the transmission of 300 Gbit/s DP-64-QAM over a 227-km-long SSMF.}
\label{fig:SNR_DP_300G}
\end{figure}

\begin{figure}[t]
\centering
\resizebox{\columnwidth}{!}{\includegraphics{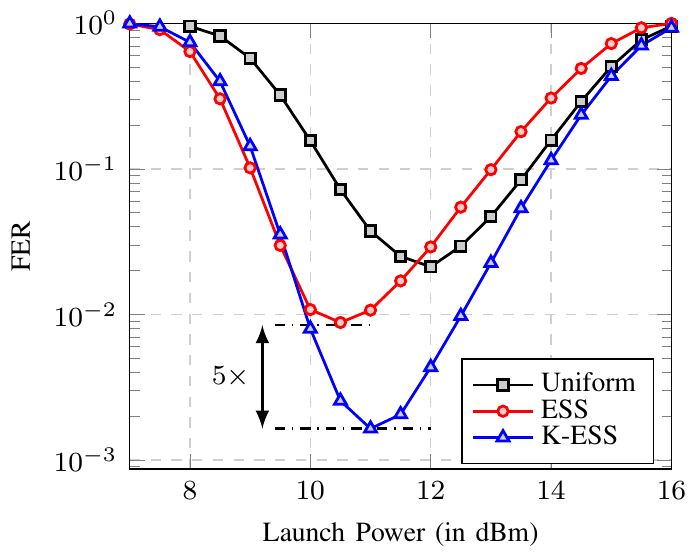}}
\caption{FER vs. launch power for the transmission of 300 Gbit/s DP-64-QAM over a 227-km-long SSMF.}
\label{fig:FER_DP_300G}
\end{figure}

Figures~\ref{fig:SNR_DP_300G} and~\ref{fig:FER_DP_300G} show the effective SNR and FER as a function of launch power for the 300 Gbit/s transmission, resp.
We see that the improvements provided by K-ESS over ESS in effective SNR (0.43 dB) and FER (5 times) at the optimum launch power are larger than their counterparts in the 400 Gbit/s case (see Figs.~\ref{fig:SNR_DP_400G} and~\ref{fig:FER_DP_400G}).
This is because at this shaping rate of 1 bit/amplitude ($H(X)\approx 2$~bit/1D symbol) the decrease in kurtosis obtained using KLSS instead of regular sphere shaping is larger than that of 1.5 bit/amplitude ($H(X)\approx 2.5$~bit/1D symbol), as seen in Fig.~\ref{fig:EssKurtosis}.
Therefore, we claim that the performance of K-ESS relative to ESS can be predicted considering the corresponding values of kurtosis, and Fig.~\ref{fig:EssKurtosis} suggests that the largest improvement can be expected for shaping rates between 0.5 and 1.5 bit/amp. for 64-QAM.

\section{Investigation of Reach Increase}\label{sec:reachincrease}
\begin{figure}[t]
\centering
\resizebox{\columnwidth}{!}{\includegraphics{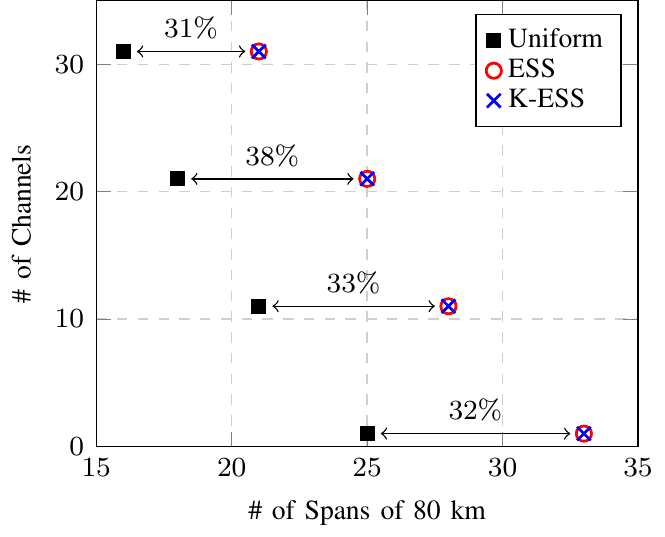}}
    \caption{Reach increase obtained using ESS and K-ESS for long-haul transmission.}
    \label{fig:longhaul}
\end{figure}

\begin{figure}[t]
\centering
\resizebox{\columnwidth}{!}{\includegraphics{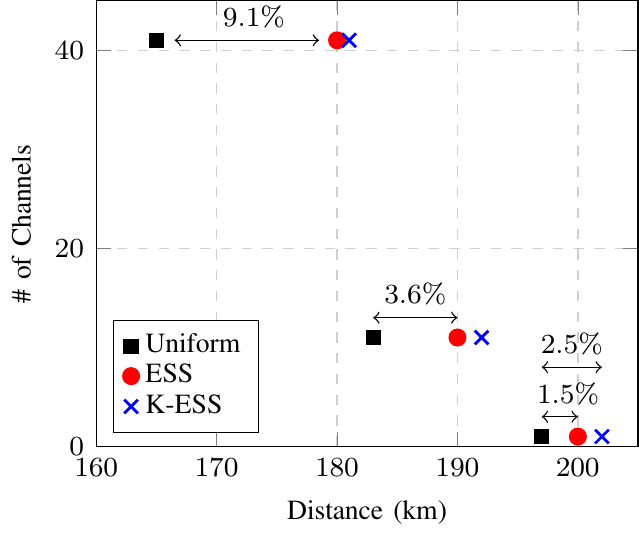}}
    \caption{Reach increase obtained using ESS and K-ESS for single-span transmission.}
    \label{fig:shorthaul}
\end{figure}

Figures~\ref{fig:longhaul} and~\ref{fig:shorthaul} show the maximum distance at which a FER of $10^{-3}$ can be achieved when using a slightly different set of parameters than that of Sec.~\ref{ssec:simulationsetup}: the attenuation is 0.19 dB/km instead of 0.2, the noise figure is 5.5 dB instead of 5, and the symbol rate is 56 GBd instead of 50~\cite{gultekin_kess_ecoc_arxiv_letter}.
When WDM is employed to realize multi-channel transmission, the channel spacing is 62.5 GHz.
This set of parameters were used in~\cite{Skvortcov2020_NLtolerantLUTshaping_letter}, and we report here that when the same link setups are simulated, we obtained results that are in agreement with~\cite{Skvortcov2020_NLtolerantLUTshaping_letter}.

In Fig.~\ref{fig:longhaul}, we consider a multi-span(s of 80 km) link for WDM transmission with $\{1, 11, 21, 31\}$ channels.
Similar to the observations of~\cite{Buchali2016_RateAdaptReachIncrease_letter,Amari2019_IntroducingESSoptics_letter}, more than 30\% reach increase is obtained for all long-haul systems under consideration using an AWGN-optimal shaping scheme, i.e., ESS.
On the other hand, it is not possible to attain a further increase using K-ESS.
Moreover, the optimum K-ESS trellis is the one that has an inactive kurtosis constraint, which is equivalent to ESS.
Therefore, we claim that AWGN-optimal shaping is good enough for multi-span systems, and modifications concerning kurtosis do not provide additional improvement.
This observation agrees with~\cite{FehenbergerABH2016_OnPSofQAMforNLFC_short}, where no improvement in AIR was secured with an EGN-model-based optimization of the input.

In Fig.~\ref{fig:shorthaul}, we consider a single-span link for WDM transmission with $\{1, 11, 41\}$ channels.
It can be seen that the largest reach increase provided by K-ESS over ESS is obtained for single-channel transmission.
In this case, the reach increase by ESS and K-ESS over uniform signaling are 1.5\% and 2.5\% resp.
As the number of channels increases to 11 and then to 41, two observations are made: (i) the reach increase by ESS also increases to 3.6\% and then to 9.1\%, and (ii) the performance of K-ESS converges to that of ESS.
Therefore, we claim that for single-span systems, as the number of channels increases, AWGN-optimal shaping is again good enough.
This observation agrees with~\cite{Renner2017_ExperimentalPScomparison_letter}, where only marginal improvements in SNR and AIR were discovered after an EGN-model-based optimization of the input for single-span transmission with 9 channels.

\section{Blocklength Dependency \& Discussion}\label{sec:discussion}
It is reported in the literature that for nonlinear optical channels, there is a finite optimum shaping blocklength in the sense that effective SNR (or the AIR) is maximized due to an increased tolerance to NLI~\cite{Amari2019_IntroducingESSoptics_letter,Goossens2019_FirstExperimentESS_letter,Fehenberger2019_NLFiberInteractions_letter,Fehenberger2020_MitigatingNLbyShortShaping,Civelli2020_interplayPSandCPR,wu2021_arxiv_EDI,Skvortcov2021_HCSSforExtendedReachSingleSpanLins_letter}.
This optimum can be, e.g., around a few dozens of amplitudes for single-span transmission with 9 WDM channels~\cite[Fig. 11(a)]{Skvortcov2021_HCSSforExtendedReachSingleSpanLins_letter}, or around a few hundreds of amplitudes for 10-span transmission with 5 WDM channels~\cite[Fig. 3]{Fehenberger2020_MitigatingNLbyShortShaping}.
These observations create a prima facie assumption that the optimum blocklength depends on the channel memory which depends on the transmission distance and the bandwidth~\cite[eq. (2)]{wu2021_arxiv_EDI}.
On the contrary, authors of~\cite{Civelli2020_interplayPSandCPR} observed that when a carrier phase recovery algorithm is employed to mitigate laser phase noise, this behavior disappears (for high SNR) and AIR converges to its maximum as $N$ increases.
For the simulations discussed in Secs.~\ref{sec:numeric} and~\ref{sec:reachincrease}, we selected $N=108$ as an average value expected to create a small rate loss without resorting to an optimization over $N$.
Now, we investigate how the performance of K-ESS depends on shaping blocklength.

\begin{figure}[t]
\centering
\resizebox{\columnwidth}{!}{\includegraphics{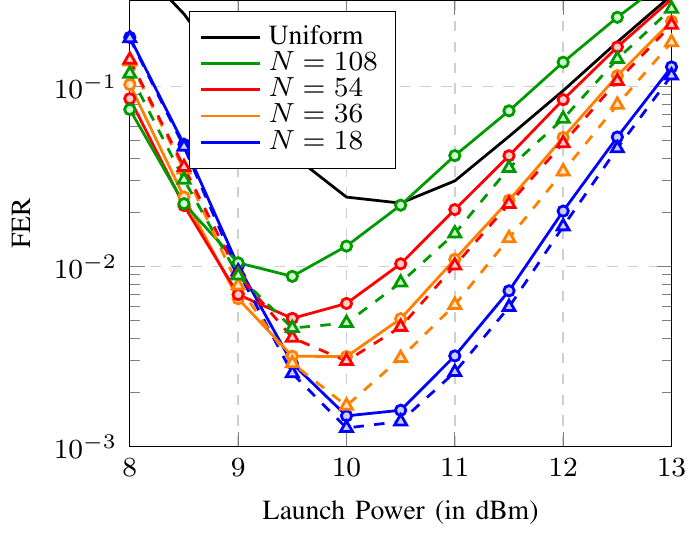}}
\caption{The transmission of 400 Gbit/s DP-64-QAM over a 205-km-long SSMF using ESS (solid-circle) and K-ESS (dashed-triangle)  at various blocklengths.}
\label{fig:dependence_blocklength_DP400}
\end{figure}

Figure~\ref{fig:dependence_blocklength_DP400} shows the FER for DP transmission for various shaping blocklengths.
We note that this figure is an extended version of Fig.~\ref{fig:FER_DP_400G}: in addition to $N=108$, we now consider $N \in \{54, 36, 18\}$.
There are two main observations.
Firstly, $N=108$ is not the optimum blocklength, and it is possible to further decrease the FERs by decreasing $N$ until $N=18$.
For $N<18$, FERs start to increase again, and thus, we did not include them in Fig.~\ref{fig:dependence_blocklength_DP400} for the sake of clarity.
Secondly, although K-ESS consistently performs better than ESS, the gap between them diminishes as we approach the optimum blocklength.
We believe there are several reasons behind this behavior.
(i) As $N$ decreases, $\mu_4$ of sphere shaping decreases (see Fig.~\ref{fig:EssKurtosis}) limiting the room for improvement that can be exploited by KLSS.
(ii) As $N$ decreases, the number of possible $(\emax,\kmax)$ pairs decreases (i.e., the corresponding contour line of constant rate shrinks, see Fig.~\ref{fig:histogramKurtosis}) narrowing the optimization space down.
(iii) As $N$ decreases, temporal structures of the sequences in the shaping sets of ESS and K-ESS become similar.
This can also be seen by considering the extreme when $N=1$ where the difference between ESS and K-ESS disappears, and they both output the same i.i.d. amplitudes.

The above three explanations can be considered as different interpretations of the same phenomenon.
However, the third argument (iii) brings a concept that we did not discuss before in this paper, {\it temporal shaping}, into the discussion.
There has been some recent research on temporal shaping aspects for optical channels.
The investigation of the shaping gain and sphere shaping for the fiber channel in~\cite{Dar2014_ShapingGainforFOchan} is considered to be among the earliest studies on temporal shaping by both~\cite{Yankov2017_TemporalPS_jlt} and~\cite{wu2021_arxiv_EDI}.
In~\cite{Yankov2017_TemporalPS_jlt}, the state transition probabilities of a finite state machine source were optimized to realize temporal shaping over a few time slots for the nonlinear fiber channel.
In~\cite{Fehenberger2019_NLFiberInteractions_letter}, temporal structures imposed by short-blocklength shaping (e.g., unlikely long runs of identical symbols) were considered to improve the tolerance to NLI.
Recently in~\cite{wu2021_arxiv_ListCCDM_letter}, temporal shaping was studied based on constant composition sequences.
More recently in~\cite{Civelli2021_SequenceSelectionBound}, AIRs were computed for optical communications based on the selection and transmission of sequences with {\it good} temporal properties.
However, it would be misleading to consider only these works as references for temporal shaping.
In fact, any finite-blocklength amplitude shaper realizes some sort of temporal shaping.
As an example, CCDM~\cite{SchulteB2016_CCDM} and the associated MPDM~\cite{Fehenberger2019_MPDM} put a constraint on the composition of amplitudes in each sequence and hence, create dependency over multiple time slots.
Similarly, all sphere shaping implementations~\cite{Geller2016_ShapingNLphaseNoise,Schulte2019_SMDM,GultekinHKW2019_ESSforShortWlessComm,Millar2019ECOC_HCSS_letter} put a constraint on the energy of the sequences and hence, they too create dependency.
Therefore, practical amplitude shapers realize temporal shaping with correlation between channel inputs, rather than symbol-by-symbol shaping with i.i.d. input sequences.
The nuance here is the following.
The NLI generation is almost always studied for asymptotically long channel inputs (one of the exceptions being~\cite{Agrell2014_FiniteMemoryGN}).
Moreover, in theory, almost all existing shaping approaches start from an optimization objective (e.g., average distribution, average energy, kurtosis, etc.) that does not prescribe specific temporal structures. 
However, in practice, a finite blocklength must be selected to realize the corresponding shaper, which indirectly imposes such temporal structures.
The study of ``good'' temporal characteristics for the optical channel and shaping architectures designed specifically to obtain such characteristics, therefore, seem to be the natural next step of this research area.

To wrap up the above discussion, although the gains provided by KLSS are smaller around the optimum $N$ for the set of parameters considered in Fig.~\ref{fig:dependence_blocklength_DP400}, we believe that the K-ESS algorithm introduced in this paper can still find an area of use for mainly three reasons.
First: As demonstrated in Sec.~\ref{ssec:rateadapt}, gains obtained with K-ESS are larger for a 6 bit/4D transmission than that of 8 bit/4D, and expected to follow a similar behavior even after an optimization over $N$.
Since ESS can be recovered as a special case of K-ESS as we discussed in Sec.~\ref{sec:intro} (i.e., by turning off the kurtosis constraint), the optimum performance can be obtained through K-ESS at any rate.
Consequently, K-ESS can be a valuable tool in the context of elastic optical networks where rate adaptation is crucial.
Second: It can be deduced from the above discussion and from Figs.~\ref{fig:longhaul},~\ref{fig:shorthaul}, and~\ref{fig:dependence_blocklength_DP400} that the advantages of KLSS become apparent for single-span transmission with a few WDM channels when the optimum $N$ is relatively large.
Assuming that the optimum $N$ indeed follows a similar trend as the channel memory, K-ESS can be more beneficial for another set of parameters which leads to a larger memory, e.g., a higher bandwidth.
Moreover, there exist systems for which the performance improves as $N$ increases, e.g., the ones employing a carrier phase recovery algorithm~\cite[Fig. 2]{Civelli2020_interplayPSandCPR}.
Third: It is possible to generalize the shaping set in~\eqref{eq:EKlimitedSet} such that constraints on moments other than the second and the fourth (or even more than two constraints) are realized.
As we discussed in footnote 2, the sixth order moment of the channel input also plays a role in NLI generation.
Thus, a shaping set which provides a better performance can be found and realized with K-ESS.
This generalization can even be used to improve tolerance to nonlinearities caused by other phenomena in communication systems such as power amplifiers.
However, we leave for future research the further investigation and generalization of K-ESS for systems with larger optimum $N$, with a carrier phase recovery algorithm, or with other sources of nonlinearities.

\section{Conclusions}\label{sec:conclusion}
We proposed kurtosis-limited sphere shaping (KLSS) as a constellation shaping technique tailored for optical channels.
KLSS creates channel input distributions with a smaller fourth order moment (i.e., kurtosis) than that of Gaussian-channel-optimized inputs.
KLSS is implemented using a modified version of the enumerative sphere shaping algorithm.
Simulations of transmission over SSMF demonstrate that KLSS can recover most of the effective SNR-penalty experienced with Gaussian-channel-optimized inputs.
End-to-end decoding results show that smaller frame error probabilities can be achieved with KLSS.
Future work should focus on further generalizations of KLSS for other higher order moments and on shaping strategies that optimize the temporal structure of the channel inputs for nonlinear channels with memory.

\ifCLASSOPTIONcaptionsoff
  \newpage
\fi

\bibliographystyle{IEEEtran}
\bibliography{IEEEabrv,PhD_refs}

\end{document}

%% file: EmaxKmaxPlane_Histograms.tikz
\begin{tikzpicture}
\begin{axis}[name=outset,
every axis/.append style={font=\footnotesize},
width=\columnwidth,
xmin=1.3,
xmax=3.2,
ymin=0,
ymax=0.2,
xlabel={Kurtosis},
ylabel={Probability},
ylabel near ticks,
xlabel near ticks,
legend style={at={(0.715,0.235)},anchor=center,font=\footnotesize,legend cell align=left,row sep=-0.5ex},
area style]

\node[anchor=west] (p1) at (axis cs:1.8554, 0.095){};
\node[anchor=west] (p2) at (axis cs:1.8554, 0.125){};
\draw[-latex, semithick, color=red] (p1)--(p2) node[above, align=center, pos=0.92] {Mean \\ 1.86};

\node[anchor=west] (p3) at (axis cs:1.6508, 0.18){};
\node[anchor=west] (p4) at (axis cs:1.6508, 0.15){};
\draw[-latex, semithick, color=blue] (p4)--(p3) node[above, align=center, pos=0.92] {Mean \\ 1.65};

\node[anchor=west] (p5) at (axis cs:1.47, 0.175){};
\node[anchor=west] (p6) at (axis cs:1.47, 0.145){};
\draw[-latex, semithick, color=black] (p6)--(p5) node[above, align=center, pos=0.92] {Mean \\ 1.58};

\addplot+[ybar interval,mark=no, black, opacity=0.45]
  table[]{%
1	2.52751910754755e-26
1.02914572864322	1.01254969675157e-21
1.05829145728643	2.29422703063009e-18
1.08743718592965	3.46792318747849e-16
1.11658291457286	2.46050254283791e-14
1.14572864321608	1.00704050879838e-12
1.17487437185930	1.28203517398461e-10
1.20402010050251	8.33580009797495e-09
1.23316582914573	1.04294089995336e-06
1.26231155778894	1.55589302404055e-05
1.29145728643216	0.000119372090169460
1.32060301507538	0.000647350233705603
1.34974874371859	0.00709777886663528
1.37889447236181	0.0196502910548830
1.40804020100503	0.0197882157328286
1.43718592964824	0.0481262549364642
1.46633165829146	0.0779469125886682
1.49547738693467	0.146334655711723
1.52462311557789	0.0695721101228612
1.55376884422111	0.0892009702665507
1.58291457286432	0.116050756336242
1.61206030150754	0.0799768504714064
1.64120603015075	0.0671087135583387
1.67035175879397	0.0823930013605154
1.69949748743719	0.0371343438872526
1.72864321608040	0.0272074660627821
1.75778894472362	0.0414613260676369
1.78693467336683	0.0266461292893188
1.81608040201005	0.0125540629685436
1.84522613065327	0.0112170385405491
1.87437185929648	0.00635187399330105
1.90351758793970	0.00167417775254367
1.93266331658291	0.00317014691905230
1.96180904522613	0.00144219919842505
1.99095477386935	0.00326442464786245
2.02010050251256	0.00145627369939945
2.04924623115578	0.00119416372633994
2.07839195979900	0.000235839660562940
2.10753768844221	0.000385349930345301
2.13668341708543	0.000269812783719599
2.16582914572864	8.18568354455990e-05
2.19497487437186	8.58620404672407e-05
2.22412060301508	2.57099230069150e-05
2.25326633165829	5.83558255325012e-05
2.28241206030151	1.08392014484452e-05
2.31155778894472	2.25716757572452e-05
2.34070351758794	7.50428323196041e-06
2.36984924623116	5.13983983222535e-06
2.39899497487437	1.61450770527755e-06
2.42814070351759	3.15970687790639e-06
2.45728643216080	4.50354158328486e-07
2.48643216080402	1.14059735429402e-06
2.51557788944724	2.76832101989465e-07
2.54472361809045	4.25712460545675e-07
2.57386934673367	4.57138199605318e-08
2.60301507537688	7.97663136493616e-08
2.63216080402010	3.02243619903654e-07
2.66130653266332	2.30358836706906e-08
2.69045226130653	5.52229842247822e-08
2.71959798994975	6.09284135236806e-08
2.74874371859297	2.90349827221661e-09
2.77788944723618	3.46193451651941e-08
2.80703517587940	8.27565088927965e-10
2.83618090452261	2.22329390079607e-09
2.86532663316583	8.14524770003791e-09
2.89447236180905	4.05189235075996e-10
2.92361809045226	7.20784197261134e-10
2.95276381909548	2.05115065493702e-09
2.98190954773869	6.11089372163878e-11
3.01105527638191	7.77163885895873e-11
3.04020100502513	6.98715524632465e-10
3.06934673366834	1.16308536329460e-11
3.09849246231156	4.92064959798257e-12
3.12763819095477	1.16092942295638e-10
3.15678391959799	4.79862934760004e-11
3.18592964824121	2.05298160292155e-12
3.21507537688442	7.70166331630959e-13
3.24422110552764	3.28034309473070e-11
3.27336683417085	4.46843905231363e-15
3.30251256281407	1.00324776054287e-12
3.33165829145729	1.15693459267725e-13
3.36080402010050	6.10712046808035e-12
3.38994974874372	1.36836543621876e-13
3.41909547738693	1.19286259347706e-13
3.44824120603015	4.17054311549272e-14
3.47738693467337	1.09314619144690e-12
3.50653266331658	1.86895985983631e-16
3.53567839195980	2.63141938761955e-14
3.56482412060302	1.78737562092545e-14
3.59396984924623	1.32989973964953e-14
3.62311557788945	1.66268059475057e-13
3.65226130653266	2.65580324961097e-15
3.68140703517588	4.48348430164802e-15
3.71055276381910	2.20555832694614e-14
3.73969849246231	1.02056941318274e-15
3.76884422110553	2.53221048560510e-14
3.79798994974874	9.02714960063360e-18
3.82713567839196	4.96498689105387e-16
3.85628140703518	3.64095033945249e-15
3.88542713567839	2.87179496066940e-16
3.91457286432161	3.48670448196169e-15
3.94371859296482	2.73053526939915e-20
3.97286432160804	0
4.00201005025126	4.69702126149927e-16
4.03115577889447	0
4.06030150753769	1.37967575913777e-16
4.08944723618090	3.91139165378774e-16
4.11859296482412	0
4.14773869346734	0
4.17688442211055	4.21799512874199e-17
4.20603015075377	1.20361994675115e-17
4.23517587939698	7.28142738506441e-20
4.26432160804020	2.72398198475260e-17
4.29346733668342	2.81390761295814e-17
4.32261306532663	1.85121035213502e-21
4.35175879396985	3.78861768629133e-18
4.38090452261307	0
4.41005025125628	0
4.43919597989950	1.55913563882692e-18
4.46834170854271	1.06126804137314e-18
4.49748743718593	3.88652357378987e-18
4.52663316582915	0
4.55577889447236	2.63951742708585e-19
4.58492462311558	3.08535058689170e-23
4.61407035175879	0
4.64321608040201	9.25605176067509e-22
4.67236180904523	1.85121035213502e-21
4.70150753768844	4.77396296309753e-19
4.73065326633166	5.46107053879831e-20
4.75979899497487	1.64294918751983e-20
4.78894472361809	5.55363105640506e-22
4.81809045226131	0
4.84723618090452	0
4.87638190954774	0
4.90552763819095	0
4.93467336683417	1.85121035213502e-21
4.96381909547739	1.86663710506948e-20
4.99296482412060	1.37452368646025e-20
5.02211055276382	2.82309578700590e-21
5.05125628140704	0
5.08040201005025	0
5.10954773869347	0
5.13869346733668	1.51738553453690e-24
5.16783919597990	3.08535058689170e-23
5.19698492462312	3.08535058689170e-23
5.22613065326633	9.33318552534739e-22
5.25527638190955	0
5.28442211055276	0
5.31356783919598	9.25605176067509e-22
5.34271356783920	9.25605176067510e-23
5.37185929648241	0
5.40100502512563	3.03477106907380e-24
5.43015075376884	0
5.45929648241206	0
5.48844221105528	0
5.51758793969849	0
5.54673366834171	3.08535058689170e-23
5.57587939698492	0
5.60502512562814	0
5.63417085427136	1.51738553453690e-24
5.66331658291457	0
5.69246231155779	4.77976443379124e-23
5.72160804020101	0
5.75075376884422	0
5.77989949748744	0
5.80904522613065	0
5.83819095477387	4.89479204689323e-26
5.86733668341709	5.05795178178967e-25
5.89648241206030	0
5.92562814070352	0
5.95477386934673	0
5.98391959798995	0
6.01306532663317	0
6.04221105527638	0
6.07135678391960	0
6.10050251256281	1.51738553453690e-24
6.12964824120603	0
6.15879396984925	4.89479204689323e-26
6.18793969849246	0
6.21708542713568	0
6.24623115577889	0
6.27537688442211	0
6.30452261306533	0
6.33366834170854	0
6.36281407035176	0
6.39195979899498	0
6.42110552763819	0
6.45025125628141	0
6.47939698492462	0
6.50854271356784	0
6.53768844221106	0
6.56683417085427	2.44739602344661e-26
6.59597989949749	0
6.62512562814070	0
6.65427135678392	0
6.68341708542714	0
6.71256281407035	0
6.74170854271357	0
6.77085427135678	0
6.80000000000000	7.76951118554481e-28
};
  \addlegendentry{$\kmax=10320$};
  
 \addplot+[ybar interval,mark=no, blue, opacity=0.45]
  table[]{%
1	1.24687894203470e-29
1.02914572864322	1.03834991026567e-21
1.05829145728643	2.35484983976987e-18
1.08743718592965	3.50793064597771e-16
1.11658291457286	2.29493482031461e-14
1.14572864321608	6.29909034668265e-13
1.17487437185930	1.52439586833657e-11
1.20402010050251	3.30336585020564e-10
1.23316582914573	8.61573079214031e-09
1.26231155778894	1.56232097763459e-07
1.29145728643216	1.94140296577898e-06
1.32060301507538	8.70000225226265e-06
1.34974874371859	8.26431963507142e-05
1.37889447236181	0.000542914949812217
1.40804020100503	0.00343145357048689
1.43718592964824	0.00418188973820481
1.46633165829146	0.0179794043045901
1.49547738693467	0.0445515832668338
1.52462311557789	0.0334533072719821
1.55376884422111	0.0916178606582971
1.58291457286432	0.119195138702329
1.61206030150754	0.0821438143607999
1.64120603015075	0.152760032083849
1.67035175879397	0.151164116442981
1.69949748743719	0.0878232816307567
1.72864321608040	0.0628417106290090
1.75778894472362	0.0425847160969437
1.78693467336683	0.0273681032057924
1.81608040201005	0.0128942139116934
1.84522613065327	0.0252302490229378
1.87437185929648	0.0154587878491837
1.90351758793970	0.00748650067891343
1.93266331658291	0.00688262772059655
1.96180904522613	0.00148127542568179
1.99095477386935	0.00335287387147989
2.02010050251256	0.00149573127369840
2.04924623115578	0.00122651946000224
2.07839195979900	0.00112494182938439
2.10753768844221	0.000792597739946383
2.13668341708543	0.000277123331156400
2.16582914572864	0.000251785644461475
2.19497487437186	8.81884629265581e-05
2.22412060301508	9.39412697844269e-05
2.25326633165829	5.99369701502164e-05
2.28241206030151	1.11328884089868e-05
2.31155778894472	2.31832528073635e-05
2.34070351758794	7.70761095346469e-06
2.36984924623116	1.41441127657110e-05
2.39899497487437	1.65825261240833e-06
2.42814070351759	7.02727568861439e-06
2.45728643216080	4.62556454277674e-07
2.48643216080402	1.17150171304939e-06
2.51557788944724	1.79438369718391e-06
2.54472361809045	4.37247092427649e-07
2.57386934673367	6.05426172562876e-07
2.60301507537688	8.19275730669227e-08
2.63216080402010	3.10432877248377e-07
2.66130653266332	2.36600383827157e-08
2.69045226130653	5.67192448548808e-08
2.71959798994975	6.25792621275658e-08
2.74874371859297	2.98216823573386e-09
2.77788944723618	4.72644467289511e-08
2.80703517587940	8.49987873186902e-10
2.83618090452261	2.28353380234429e-09
2.86532663316583	1.08049151302766e-08
2.89447236180905	4.16167792441100e-10
2.92361809045226	7.40313765108630e-10
2.95276381909548	2.55470086533583e-09
2.98190954773869	6.27646770896961e-11
3.01105527638191	7.98221055151109e-11
3.04020100502513	7.17647144243799e-10
3.06934673366834	1.19459903215860e-11
3.09849246231156	5.05397405285091e-12
3.12763819095477	1.19238467685614e-10
3.15678391959799	4.92864767732347e-11
3.18592964824121	2.10860691165649e-12
3.21507537688442	7.91033902929826e-13
3.24422110552764	4.48641450377610e-11
3.27336683417085	4.58951091262378e-15
3.30251256281407	1.03043064729569e-12
3.33165829145729	1.18828160709385e-13
3.36080402010050	8.99234874866450e-12
3.38994974874372	1.40544114588103e-13
3.41909547738693	1.22518307308881e-13
3.44824120603015	4.28354351844886e-14
3.47738693467337	1.72023728667058e-12
3.50653266331658	1.91959911986117e-16
3.53567839195980	2.70271740394816e-14
3.56482412060302	1.83580436504951e-14
3.59396984924623	1.30779942049785e-13
3.62311557788945	1.70773074097642e-13
3.65226130653266	2.72776194397464e-15
3.68140703517588	4.60496381131948e-15
3.71055276381910	2.26531768508881e-14
3.73969849246231	1.04822162819185e-15
3.76884422110553	2.60082044774173e-14
3.79798994974874	9.27173921742177e-18
3.82713567839196	5.09951265995836e-16
3.85628140703518	3.73960148490475e-15
3.88542713567839	2.94960592647804e-16
3.91457286432161	3.58117635302016e-15
3.94371859296482	2.80451881954681e-20
3.97286432160804	2.16340581739841e-16
4.00201005025126	4.82428653140411e-16
4.03115577889447	0
4.06030150753769	1.41705792074455e-16
4.08944723618090	4.01737037664410e-16
4.11859296482412	0
4.14773869346734	1.54528986957030e-17
4.17688442211055	4.33228123873190e-17
4.20603015075377	1.23623189565624e-17
4.23517587939698	7.47871685212484e-20
4.26432160804020	2.79778797437990e-17
4.29346733668342	2.89015012750364e-17
4.32261306532663	8.15211826359794e-19
4.35175879396985	3.89126986212121e-18
4.38090452261307	0
4.41005025125628	0
4.43919597989950	1.60138024596123e-18
4.46834170854271	1.09002298119720e-18
4.49748743718593	4.01987368697348e-18
4.52663316582915	0
4.55577889447236	2.71103485889525e-19
4.58492462311558	3.16894781869697e-23
4.61407035175879	0
4.64321608040201	9.50684345609090e-22
4.67236180904523	1.90136869121818e-21
4.70150753768844	4.90331295986981e-19
4.73065326633166	5.60903763909363e-20
4.75979899497487	1.68746471345613e-20
4.78894472361809	5.70410607365454e-22
4.81809045226131	0
4.84723618090452	0
4.87638190954774	0
4.90552763819095	0
4.93467336683417	1.90136869121818e-21
4.96381909547739	1.91721343031166e-20
4.99296482412060	1.41176625322950e-20
5.02211055276382	2.89958725410772e-21
5.05125628140704	0
5.08040201005025	0
5.10954773869347	0
5.13869346733668	1.55849892722802e-24
5.16783919597990	3.16894781869697e-23
5.19698492462312	3.16894781869697e-23
5.22613065326633	9.58606715155832e-22
5.25527638190955	0
5.28442211055276	0
5.31356783919598	9.50684345609090e-22
5.34271356783920	9.50684345609090e-23
5.37185929648241	0
5.40100502512563	3.11699785445603e-24
5.43015075376884	0
5.45929648241206	0
5.48844221105528	0
5.51758793969849	0
5.54673366834171	3.16894781869697e-23
5.57587939698492	0
5.60502512562814	0
5.63417085427136	1.55849892722802e-24
5.66331658291457	0
5.69246231155779	4.90927162076825e-23
5.72160804020101	0
5.75075376884422	0
5.77989949748744	0
5.80904522613065	0
5.83819095477387	5.02741589428392e-26
5.86733668341709	5.19499642409339e-25
5.89648241206030	0
5.92562814070352	0
5.95477386934673	0
5.98391959798995	0
6.01306532663317	0
6.04221105527638	0
6.07135678391960	0
6.10050251256281	1.55849892722802e-24
6.12964824120603	0
6.15879396984925	5.02741589428392e-26
6.18793969849246	0
6.21708542713568	0
6.24623115577889	0
6.27537688442211	0
6.30452261306533	0
6.33366834170854	0
6.36281407035176	0
6.39195979899498	0
6.42110552763819	0
6.45025125628141	0
6.47939698492462	0
6.50854271356784	0
6.53768844221106	0
6.56683417085427	2.51370794714196e-26
6.59597989949749	0
6.62512562814070	0
6.65427135678392	0
6.68341708542714	0
6.71256281407035	0
6.74170854271357	0
6.77085427135678	0
6.80000000000000	7.98002522902210e-28
};
\addlegendentry{$\emax=552$, $\kmax=10624$};

\addplot+[ybar interval,mark=no, red, opacity=0.45]
  table[]{%
1	8.19469470582268e-30
1.02914572864322	6.14392056522001e-22
1.05829145728643	1.50832518457036e-18
1.08743718592965	2.22132116852740e-16
1.11658291457286	1.46916891642195e-14
1.14572864321608	3.77277120913172e-13
1.17487437185930	8.09149361095828e-12
1.20402010050251	1.82914238608342e-10
1.23316582914573	2.08592260176562e-09
1.26231155778894	3.52716208733406e-08
1.29145728643216	4.43652792464443e-07
1.32060301507538	8.95308954076411e-07
1.34974874371859	1.04152308689271e-05
1.37889447236181	0.000117959660374462
1.40804020100503	0.000190504204737876
1.43718592964824	0.00100382630505980
1.46633165829146	0.00416894665985871
1.49547738693467	0.00374183150471525
1.52462311557789	0.00959163100930270
1.55376884422111	0.00900319141521862
1.58291457286432	0.0274960133950872
1.61206030150754	0.0226876348910901
1.64120603015075	0.0180533041330434
1.67035175879397	0.0556172331295829
1.69949748743719	0.0577189137425906
1.72864321608040	0.0413006119548411
1.75778894472362	0.0604431543395480
1.78693467336683	0.0744074791438274
1.81608040201005	0.0470209018489186
1.84522613065327	0.0794537331410646
1.87437185929648	0.0942513402682626
1.90351758793970	0.0476432214146470
1.93266331658291	0.0650430441617644
1.96180904522613	0.0564156225752316
1.99095477386935	0.0381664427341697
2.02010050251256	0.0420026704075867
2.04924623115578	0.0454189549579430
2.07839195979900	0.0248003350136828
2.10753768844221	0.0151002097247022
2.13668341708543	0.0206276711678630
2.16582914572864	0.00772711823662771
2.19497487437186	0.0110830069061902
2.22412060301508	0.00895999353259798
2.25326633165829	0.000981802580049333
2.28241206030151	0.00430599385649860
2.31155778894472	0.00283830055186479
2.34070351758794	0.000106771857133924
2.36984924623116	0.00143971094836413
2.39899497487437	0.000232556196282566
2.42814070351759	0.000397820040435294
2.45728643216080	0.000207278505737855
2.48643216080402	7.60716614910941e-05
2.51557788944724	7.30208020203819e-05
2.54472361809045	3.31266437360883e-05
2.57386934673367	1.72959171768452e-05
2.60301507537688	1.01077776895543e-05
2.63216080402010	5.15415345029280e-06
2.66130653266332	3.49403483369065e-06
2.69045226130653	1.11717694228177e-06
2.71959798994975	1.26941799516823e-06
2.74874371859297	2.08127001626243e-07
2.77788944723618	2.73441189862212e-07
2.80703517587940	1.83706689303934e-07
2.83618090452261	4.79417480403763e-08
2.86532663316583	6.55216235915862e-08
2.89447236180905	4.22870155906518e-09
2.92361809045226	1.49625294477116e-08
2.95276381909548	9.89925626072626e-09
2.98190954773869	7.22679305507520e-10
3.01105527638191	2.74851930048512e-09
3.04020100502513	1.97481053211134e-09
3.06934673366834	3.03260966508596e-10
3.09849246231156	3.33683868994374e-10
3.12763819095477	3.53880430641362e-10
3.15678391959799	1.01687039550178e-10
3.18592964824121	6.89407316733743e-11
3.21507537688442	1.45569573949700e-11
3.24422110552764	7.18842312641809e-11
3.27336683417085	8.54697243861422e-12
3.30251256281407	5.85643462690430e-12
3.33165829145729	7.61317382185805e-12
3.36080402010050	7.63645396367782e-12
3.38994974874372	1.06230213529915e-12
3.41909547738693	1.55305726067484e-13
3.44824120603015	1.15691958349287e-12
3.47738693467337	1.30464460675210e-12
3.50653266331658	1.38691237123184e-13
3.53567839195980	2.34378232570649e-14
3.56482412060302	1.65040193154053e-13
3.59396984924623	1.05657251404288e-13
3.62311557788945	1.33550384319577e-13
3.65226130653266	2.09080596635911e-15
3.68140703517588	7.16397234976535e-15
3.71055276381910	3.07887978150845e-14
3.73969849246231	3.50412426955964e-15
3.76884422110553	1.70933124624198e-14
3.79798994974874	2.43741616663408e-17
3.82713567839196	2.31622610470101e-15
3.85628140703518	2.77052970976535e-15
3.88542713567839	1.93858072350862e-16
3.91457286432161	2.35364398526790e-15
3.94371859296482	1.31909974535274e-18
3.97286432160804	1.63510001178370e-16
4.00201005025126	3.45496614672526e-16
4.03115577889447	0
4.06030150753769	9.31313911037745e-17
4.08944723618090	2.64063862716987e-16
4.11859296482412	7.48329524843798e-19
4.14773869346734	1.21870808331704e-17
4.17688442211055	2.84724690860832e-17
4.20603015075377	8.12472055544695e-18
4.23517587939698	4.97657565782821e-20
4.26432160804020	1.84004277007774e-17
4.29346733668342	1.89945448194342e-17
4.32261306532663	6.42674917971387e-19
4.35175879396985	2.55740693527283e-18
4.38090452261307	0
4.41005025125628	0
4.43919597989950	1.05245359282219e-18
4.46834170854271	7.16381137904653e-19
4.49748743718593	2.64561384908249e-18
4.52663316582915	0
4.55577889447236	1.78173696391380e-19
4.58492462311558	2.08268493736272e-23
4.61407035175879	0
4.64321608040201	6.24805481208815e-22
4.67236180904523	1.31209151053851e-21
4.70150753768844	3.22253840358133e-19
4.73065326633166	3.68635233913201e-20
4.75979899497487	1.10902972914565e-20
4.78894472361809	3.74883288725289e-22
4.81809045226131	0
4.84723618090452	0
4.87638190954774	0
4.90552763819095	0
4.93467336683417	1.24961096241763e-21
4.96381909547739	1.26002438710444e-20
4.99296482412060	9.27836139595090e-21
5.02211055276382	1.90565671768689e-21
5.05125628140704	0
5.08040201005025	0
5.10954773869347	0
5.13869346733668	1.02427128067019e-24
5.16783919597990	2.08268493736272e-23
5.19698492462312	2.08268493736272e-23
5.22613065326633	6.30012193552222e-22
5.25527638190955	0
5.28442211055276	0
5.31356783919598	6.24805481208815e-22
5.34271356783920	6.24805481208815e-23
5.37185929648241	0
5.40100502512563	2.04854256134038e-24
5.43015075376884	0
5.45929648241206	0
5.48844221105528	0
5.51758793969849	0
5.54673366834171	2.08268493736272e-23
5.57587939698492	0
5.60502512562814	0
5.63417085427136	1.02427128067019e-24
5.66331658291457	0
5.69246231155779	3.22645453411109e-23
5.72160804020101	0
5.75075376884422	0
5.77989949748744	0
5.80904522613065	0
5.83819095477387	3.30410090538770e-26
5.86733668341709	3.41423760223396e-25
5.89648241206030	0
5.92562814070352	0
5.95477386934673	0
5.98391959798995	0
6.01306532663317	0
6.04221105527638	0
6.07135678391960	0
6.10050251256281	1.02427128067019e-24
6.12964824120603	0
6.15879396984925	3.30410090538770e-26
6.18793969849246	0
6.21708542713568	0
6.24623115577889	0
6.27537688442211	0
6.30452261306533	0
6.33366834170854	0
6.36281407035176	0
6.39195979899498	0
6.42110552763819	0
6.45025125628141	0
6.47939698492462	0
6.50854271356784	0
6.53768844221106	0
6.56683417085427	1.65205045269385e-26
6.59597989949749	0
6.62512562814070	0
6.65427135678392	0
6.68341708542714	0
6.71256281407035	0
6.74170854271357	0
6.77085427135678	0
6.80000000000000	5.24460461172652e-28
};
  \addlegendentry{$\emax=528$};

\end{axis}

\begin{axis}[name=insetfig2,axis background/.style={fill=gray!2!white},draw,tiny,anchor=south east, at={(outset.north west)},yshift=-3.65cm,xshift=7.15cm,
width=0.67\columnwidth,
xmin=-1,
xmax=1,
ymin=-1,
ymax=1,
ticks = none,
]

\node[draw=none] at (0.03,0.02) {\scalebox{0.7}{\input{EmaxKmaxPlane_n64.tikz}}};

\end{axis}

\end{tikzpicture}